\global\def\draftcontrol{0}

   \def\versionno{ violation }
\catcode`\@=11

\expandafter\ifx\csname draftcontrol\endcsname\relax\global\def\draftcontrol{0}
\fi

{\count255=\time\divide\count255 by 60
\xdef\hourmin{\number\count255}
\multiply\count255 by-60\advance\count255 by\time
\xdef\hourmin{\hourmin:\ifnum\count255<10 0\fi\the\count255}}
\def\draftdate{\number\month/\number\day/\number\year\ \ \ \hourmin }

\newcommand\makepapertitle{\par
  \begingroup
    \renewcommand\thefootnote{\@fnsymbol\c@footnote}%
    \def\@makefnmark{\rlap{\@textsuperscript{\normalfont\@thefnmark}}}%
    \long\def\@makefntext##1{\parindent 1em\noindent
            \hb@xt@1.8em{%
                \hss\@textsuperscript{\normalfont\@thefnmark}}##1}%
     \newpage
     \global\@topnum\z@   % Prevents figures from going at top of page.
     \@makepapertitle
     \thispagestyle{empty}\@thanks
  \endgroup
  \setcounter{footnote}{0}%
  \global\let\thanks\relax
  \global\let\makepapertitle\relax
  \global\let\@makepapertitle\relax
  \global\let\@thanks\@empty
  \global\let\@author\@empty
  \global\let\@date\@empty
  \global\let\@title\@empty
  \global\let\title\relax
  \global\let\author\relax
  \global\let\date\relax
  \global\let\and\relax
  \def\version{\let\version\@version\@gobble}
}
\def\@makepapertitle{%
  \newpage
   \ifnum\draftcontrol=1 {}
   \version\versionno
   \vskip 3em%
   \else
   \hfill\hbox to 3cm {\parbox{4cm}{\@pubnum}\hss}%
   \vskip 3em%
   \fi
   \begin{center}%
   \let \footnote \thanks
     {\LARGE {\@title}}%
     \vskip 1.5em%
     {\normalsize%\large
       \lineskip .5em%
       \begin{tabular}[t]{c}%
         \@author
       \end{tabular}\par}%
     \vskip 1.5em%
     {\@bstract}%
     \end{center}%
     \vskip 1.5em
     \@date%
   \par
}

\gdef\@pubnum{}
\def\pubnum#1{%
  \gdef\@pubnum{#1}}

\gdef\@bstract{}
\def\Abstract#1{%
  \gdef\@bstract{%
   \parbox{\textwidth-0pc}{%
   \centerline{\bf Abstract}\penalty1000%
\kern.2cm%
\noindent%\abstractfont \baselineskip=12pt
\renewcommand\baselinestretch{1.0}%
{#1}}}
}

\def\ps@paper{\let\@mkboth\@gobbletwo%
     \ifnum\draftcontrol=1
    \def\@oddfoot{\hbox to \textwidth{\tiny \versionno \hfil\tiny\draftdate}%
    \hskip -\textwidth \hbox to \textwidth{\hfil\rm\thepage\hfil}}%
     \else\def\@oddfoot{\hbox to \textwidth{\hfil\rm\thepage\hfil}}
     \fi
     \let\@evenfoot\@oddfoot
}
\def\body{\clearpage
          \pagestyle{paper}
    }
\def\@version#1{\ifnum\draftcontrol=1
\typeout{}\typeout{#1}\typeout{}
\vskip3mm\centerline{\hbox{\fbox{\normalsize{\tt DRAFT -- #1 -- }
                   {\draftdate}}}}\vskip3mm
\fi}
\let\version\@version
\long\def\eqlabel#1{\ifnum\draftcontrol=1
                    \tag@false  % there are some problems with multline without this
                    \tag*{(\theequation) \hbox to -0.2cm{\hspace{0cm}\small{#1}\hss}}
                    \refstepcounter{equation}
                    \edef\@currentlabel{\theequation}
                    \ltx@label{#1}          % use old LaTeX \label instead of new definition
                    \else
                    \label{#1}
                    \fi
                    }
\let\st@bibitem\@bibitem
\let\st@lbibitem\@lbibitem
\ifnum\draftcontrol=1
  \def\@bibitem#1{%
    \st@bibitem{#1}\a@@label{#1}\ignorespaces}
  \def\@lbibitem[#1]#2{%
    \st@lbibitem[#1]{#2}\a@@label{#2}\ignorespaces}
  \def\a@@label#1{%
    \gdef\a@lab{\smash{\normalfont\small#1}}
    \ifvmode
      \if@inlabel
        \global\setbox\@labels\hbox{%
          \llap{\a@lab\let\a@lab\relax
                \kern\@totalleftmargin\kern\marginparsep}%
          \box\@labels}%
      \fi
    \fi}
\fi
\documentclass[12pt,letterpaper]{article}
\usepackage{amsmath,amssymb,array,calc,epsfig}
\usepackage[sort]{cite}
\ifnum\draftcontrol=1
\fi
\renewcommand\baselinestretch{1.25}
\renewcommand\section{\@startsection {section}{1}{\z@}%
                                   {-3.5ex \@plus -1ex \@minus -.2ex}%
                                   {2.3ex \@plus.2ex}%
                                   {\normalfont\large\bfseries}}
\renewcommand\subsection{\@startsection{subsection}{2}{\z@}%
                                   {-3.25ex\@plus -1ex \@minus -.2ex}%
                                   {1.5ex \@plus .2ex}%
                                   {\normalfont\normalsize\bfseries}}
\renewcommand\subsubsection{\@startsection{subsubsection}{3}{\z@}%
                                   {-3.25ex\@plus -1ex \@minus -.2ex}%
                                   {1.5ex \@plus .2ex}%
                                   {\normalfont\normalsize\it}}
\renewcommand\paragraph{\@startsection{paragraph}{4}{\z@}%
                                   {-3.25ex\@plus -1ex \@minus -.2ex}%
                                   {1.5ex \@plus .2ex}%
                                   {\normalfont\normalsize\bf}}

\numberwithin{equation}{section}

\def\revise#1       {\raisebox{-0em}{\rule{3pt}{1em}}%
                     \marginpar{\raisebox{.5em}{\vrule width3pt\
                     \vrule width0pt height 0pt depth0.5em
                     \hbox to 0cm{\hspace{0cm}{%
                     \parbox[t]{4em}{\raggedright\footnotesize{#1}}}\hss}}}}

\def\cala         {{\cal A}}

\def\calb         {{\cal B}}
\def\calc         {{\cal C}}
\def\cald         {{\cal D}}

\def\calg         {{\cal G}}

\def\calk         {{\cal K}}
\def\call         {{\cal L}}
\def\calm         {{\cal M}}
\def\caln         {{\cal N}}
\def\calo         {{\cal O}}

\def\calu         {{\cal U}}
\def\calv         {{\cal V}}

\def\zet          {{\mathbb Z}}

\def\del          {\partial}

\def\sqr#1#2{{\vcenter{\vbox{\hrule height.#2pt
 \hbox{\vrule width.#2pt height#1pt \kern#1pt
 \vrule width.#2pt}\hrule height.#2pt}}}}

\def\jsquare{\mathop{\mathchoice{\sqr{8}{32}}{\sqr{8}{32}}
{\sqr{6.3}{9}}{\sqr{4.5}{9}}}}

\def\a{\alpha}
\def\b{\beta}

\def\dim{{\rm dim}}
\def\T{{\rm T}}
\def\adj{{\rm adj}}

\newcommand{\beq}{\begin{equation}}
\newcommand{\eeq}{\end{equation}}
\newcommand{\beqa}{\begin{eqnarray}}
\newcommand{\eeqa}{\end{eqnarray}}
\newcommand{\beqar}{\begin{eqnarray*}}
\newcommand{\eeqar}{\end{eqnarray*}}

\newcommand{\labell}[1]{\label{#1}}
\newcommand{\reef}[1]{(\ref{#1})}
\renewcommand{\eqref}[1]{(\ref{#1})}

\newcommand{\eg}{{\it e.g.,}\ }
\newcommand{\ie}{{\it i.e.,}\ }

\newcommand{\mt}[1]{\textrm{\tiny #1}}
\newcommand{\tL}{{\tilde L}}
\newcommand{\hL}{{\hat L}}
\newcommand{\tal}{{\tilde\alpha}}
\newcommand{\ls}{\ell_s}
\newcommand{\gs}{g_s}
\newcommand{\lp}{\ell_{\mt P}}
\newcommand{\vareps}{\varepsilon}
\newcommand{\nc}{N_c}

\catcode`\@=12

\begin{document}

%%%
%%%%%% text starts here
%%%%%%%%%

\title{\bf Beyond $\eta/s = 1/4\pi$}
\pubnum{UWO-TH-08/15}

%\date{November 2012}

\author{
Alex Buchel,$ ^{1,2}$  Robert C. Myers$ ^{1}$ and Aninda Sinha$ ^{1}$\\[0.4cm]
\it $ ^1$Perimeter Institute for Theoretical Physics\\
\it Waterloo, Ontario N2L 2Y5, Canada\\[.5em]
 \it $ ^2$Department of Applied Mathematics\\
 \it University of Western Ontario\\
\it London, Ontario N6A 5B7, Canada }

\Abstract{ We use a low-energy effective description of gauge
theory/string theory duality to argue that the Kovtun-Son-Starinets
viscosity bound is generically violated in superconformal gauge
theories with non-equal central charges $c\ne a$. We present new
examples (of string theory constructions and of gauge theories)
where the bound is violated in a controllable setting. We consider
the comparison of results from AdS/CFT calculations to the QCD
plasma in the context of this discussion.}

\makepapertitle

\body

\version\versionno
\tableofcontents

\section{Introduction}

Over the past decade, the AdS/CFT correspondence \cite{juan,adscft}
has been developed to provide a powerful tool to investigate the
thermal and hydrodynamic properties for certain strongly coupled
gauge theories \cite{review1}. At the same time, recent experimental
results from the Relativistic Heavy Ion Collider (RHIC) have
revealed a new phase of nuclear matter, known as the strongly
coupled quark-gluon plasma (sQGP) \cite{shuryak}. Recently, there
has been great interest in  possible connections between these two
advances, in particular, using the AdS/CFT to gain theoretical
insight into the sQGP \cite{talks}. The primary motivation for this
possible connection is the observation that a wide variety of
holographic theories exhibit an exceptionally low ratio of shear
viscosity to entropy density $\eta/s=1/4\pi$ while the RHIC data
seems to indicate that this ratio is unusually small for the sQGP
and even seems yield roughly $\eta/s\sim1/4\pi$ \cite{rick}.

Motivated by the results from the AdS/CFT correspondence, Kovtun,
Son and Starinets (KSS) proposed a now celebrated bound for the
viscosity-to-entropy-density ratio \cite{kss}. That is, for all
fluids in nature the ratio $\eta/s$ is bounded from below:
\begin{equation}
\frac{\eta}{s}\ge \frac{\hbar}{4\pi k_\mt{B}}\,. \labell{bound}
\end{equation}
This bound certainly appears to be  satisfied by all common
substances observed in nature \cite{kss1}. Using the AdS/CFT
correspondence, the bound has been shown to be saturated in all
gauge theories in the planar limit and at infinite 't Hooft coupling
(with various gauge groups, matter content, with or without chemical
potentials for conserved $U(1)$ charges, with non-commutative
spatial directions, in external background fields) that allow for a
dual supergravity description
\cite{kss1,u1,bls,mps,u3,u4,u5,u55,hong,buchbinder}. The bound is
not saturated but it is still satisfied in all
four-dimensional\footnote{Preliminary analysis indicates that the
bound is satisfied under the same conditions in three-dimensional
conformal gauge theories \cite{3d}.} conformal gauge theories with
equal $a$ and $c$ central charges, again allowing for a string
theory dual and in the planar limit and with large but finite 't
Hooft coupling \cite{fu1,fu2}.

One may ask if the KSS bound \eqref{bound} is indeed of fundamental
importance to nature? However, the answer appears to be ``no''. It
was pointed out by \cite{vi1} that the bound is violated in a
nonrelativistic gas with increasing number of species and by
\cite{vi2,vi2a,kp}, that it can be violated in effective theories of
higher derivative gravity. Of course, the true question is whether
or not the violation occurs in quantum field theories that allow for
a consistent ultraviolet completion \cite{son}. In fact, Kats and
Petrov \cite{kp} proposed an explicit example of a gauge
theory/string theory duality where a violation of the KSS bound
occurs in a controllable setting --- see also \cite{keshav}.
However, one may easily draw into question the veracity of this
claim.

In particular, the calculations in \cite{kp} were presented in terms
of an effective five-dimensional gravity theory. However, the
proposed duality is between a gauge theory and a ten-dimensional
string theory. Thus, it seems the gravity calculations should be
performed within the full ten-dimensional string theory background
constructed to required order in $\alpha'$. Alternatively, beginning
with the ten-dimensional background, one could carefully perform the
Kaluza-Klein reduction but this would require keeping track of all
of the fields and their interactions in the effective
five-dimensional theory. For instance, the reduction may produce
scalar fields which it seems are likely to effect the calculations
at the order to which they must be performed to detect the potential
violation of the viscosity bound.

Our primary motivation for the present work was to examine in detail
the claimed violation of the viscosity bound \reef{bound} in
\cite{kp}. In fact, we are able to sharpen the arguments in terms of
an effective five-dimensional gravity dual and confirm that the KSS
bound will be violated as long as the central charges of the
conformal gauge theory satisfy a number of conditions: $c\sim a\gg1$
and $|c-a|/c\ll 1$ are required to guarantee the reliability of the
low energy effective action and then the inequality
\begin{equation}
c-a>0\,, \labell{cma}
\end{equation}
produces a violation of the KSS bound \cite{vi2,kp}.

An outline of the paper is as follows: In section \ref{dual}, we
examine in detail when an effective five-dimensional gravity dual
yields a reliable description of the superconformal gauge theory. In
section~\ref{gauge} we compute $(c-a)$ in variety four-dimensional
superconformal gauge theories. This produces new examples where we
can reliably state that the KSS bound is violated. Given these
observations, we consider the comparison of results from AdS/CFT
calculations to the sQGP in section \ref{plasma}. Finally, we
provide a concluding discussion in section~\ref{end}. Appendix
\ref{redef} elaborates on the discussion of field redefinitions in
the presence of other bulk fields, while appendix \ref{string}
provides an explicit realization of our effective AdS/CFT duality in
a stringy context where ten-dimensional supergravity plus probe
branes is a reliable approximation.

\section{Effective description of conformal gauge theory/string theory
duality}\label{dual}

According to the AdS/CFT correspondence \cite{adscft}, any
four-dimensional superconformal gauge theory will have a dual
description in terms of quantum gravity with a negative cosmological
constant in five dimensions. Now for particular cases where it is
sensible to consider the conformal gauge theory with large-$N_c$ and
strong coupling, our intuition is that the dual description is well
approximated by Einstein gravity in a five-dimensional AdS
spacetime. In this framework, higher curvature (or more broadly
higher derivative) interactions are expected to arise on general
grounds, \eg as quantum or stringy corrections to the classical
action. Hence a more refined description will be given by an
effective action where the cosmological constant and Einstein terms
are supplemented by such higher curvature corrections. Here we
consider when such an effective action approach yields a reliable
description of the superconformal gauge theory.

A key assumption in our discussion will be that:\\
{\it The effective five-dimensional gravity theory is described by a
sensible derivative expansion. That is, we expect that the higher
curvature terms are systematically suppressed by powers of the
Planck length, $\lp$.}\\
\noindent Hence we can expect the effective gravity action in five
dimensions to leading order to take the form
\begin{equation}
I=\frac{1}{2\lp^3}\int d^5x \sqrt{-g}\left[\frac{12}{\tL^2} + R
+\tL^2\left(\tal_1 R^2 + \tal_2 R_{ab}R^{ab} + \tal_3
R_{abcd}R^{abcd} \right) +\cdots\right]\,, \labell{act1}
\end{equation}
where the scale $\tL$ will correspond to the AdS curvature scale, at
leading order, and we assume that $\tL\gg\lp$. We have parameterized
the curvature squared couplings with the AdS curvature scale, as is
convenient for explicit calculations, but we expect that the
dimensionless couplings $\alpha_i \sim \lp^2/\tL^2 \ll 1$ in accord
with our assumption of a sensible derivative expansion. Further,
compared to these interactions, the six- and higher derivative
terms, which have been left implicit, are suppressed by further
powers of $\lp^2/\tL^2$. For example, an interaction of the form
$\lambda\, \tL^4\, R\, R_{abcd}R^{abcd}$ would have $\lambda \sim
\lp^4/\tL^4 \ll \tal_i$.

At this point, we note that we can simplify the form of the action
by making a field redefinition $g_{ab}\rightarrow g_{ab}+\delta
g_{ab}$ with \cite{vi2,kp}
\begin{equation}
\delta g_{ab}=\frac{8}{3}\,(5\a_1+\a_2)\,g_{ab} + \a_2\, \tL^2
R_{ab}-\frac{1}{3}\,(2\a_1+\a_2)\,\tL^2R\, g_{ab}\,,\labell{redef1}
\end{equation}
which then simplifies the action to
\begin{equation}
I=\frac{1}{2\lp^3}\int d^5x \sqrt{-g}\left[\frac{12}{L^2} + R +\a_3
L^2 R_{abcd}R^{abcd}  +\cdots\right]\,. \labell{act2}
\end{equation}
The implicit terms implied by the ellipsis all contain six or more
derivatives suppressed by at least $\lp^4/\tL^4$, as described
above. Hence, the field redefinition \reef{redef1} has succeeded in
eliminating the $R^2$ and $R_{ab}R^{ab}$ terms.\footnote{Of course,
the coefficients of these two interactions could be tuned to any
convenient values. For example, this would allow us to assemble the
curvature-squared terms to be the square of Weyl-curvature or the
Gauss-Bonnet term \cite{vi2}, either of which may be advantageous
for certain calculations.} This makes clear that, at this order, the
gravity action contains two and only two dimensionless small
parameters: $\lp/L$ and $\alpha_3$.

We return to this point after making a number of observations:
first, given the effective action \reef{act2}, we might consider
making a further field redefinition of the form
\begin{equation}
\delta g_{ab}=\lambda_1\,L^4\, R_{acde}R_b{}^{cde}+\lambda_2\,L^4\,
g_{ab}\, R_{cdef}R^{cdef}\,,\labell{redef12}
\end{equation}
which would modify the action by adding terms of the form
 \beqa
\delta I&=&\frac{1}{2\lp^3}\int d^5x
\sqrt{-g}\left[6L^2(\lambda_1+5\lambda_2)R_{abcd}R^{abcd}
 \vphantom{\frac{L^4}{2}}\right.
 \labell{delact2}\\
&&\qquad\qquad\left.-L^4\,\lambda_1\,R^{ab}R_{acde}R_b{}^{cde}
+\frac{L^4}{2}(\lambda_1+3\lambda_2)R R_{abcd}R^{abcd} \right]\,.
 \nonumber
 \eeqa
Hence, given the first term above, it would seem that we can use
these field redefinitions to remove the $\alpha_3$ term in
\reef{act2}. Note that the latter would require that
$\lambda_{1,2}\sim\lp^2/L^2$ and hence the six-derivative terms,
appearing in the second line of \reef{delact2}, would only be
suppressed by this same factor $\lp^2/L^2$. However, our assumption
is that the derivative expansion organizes the effective action so
that any such term is suppressed by a factor of $\lp^4/L^4$. Hence
if we wish to maintain this structure, then we must require that
$\lambda_{1,2}\sim\lp^4/L^4$ and so this field redefinition could
only make higher order corrections to $\alpha_3$.

Next, we observe that with the original field redefinitions
\reef{redef1}, Newton's constant (\ie the coefficient of the
Einstein term) has been kept fixed but the curvature scale $L$ has to
be redefined as
\begin{equation}
L^2=\tL^2\left(1-\frac{20}{3}(5\tal_1+\tal_2)+\cdots\right)\,.
\labell{redef2}
\end{equation}
In principle, the coupling $\tal_3$ was also corrected with
$\a_3=\tal_3+O(\tal_1^2,\tal_2^2,\tal_1\tal_2)$. However, we do not
specify the latter in detail, as it actually requires specifying the
field redefinition \reef{redef1} more precisely, \ie to order
$\tal_i^2$. But these expressions do illustrate the point that in
general the parameters in this effective action \reef{act2} may be
complicated functions of the microscopic parameters of the quantum
gravity theory. For example, in a string or M-theory framework, they
would arise upon the Kaluza-Klein compactification of the higher
dimensional geometry and these low energy parameters would depend on
all of the details for the compactification. In general, we would
also expect that these parameters also receive quantum
`corrections', which might in turn include both perturbative and
nonperturbative contributions.

Note, however, that the dual theory is assumed to be dual to a
four-dimensional conformal field theory. Hence at any order in the
derivative expansion, the gravity theory admits a five-dimensional
anti-de Sitter vacuum, although the precise characteristics, \ie
curvature, of the latter may change as we increase the accuracy of
our calculations. Given the action \reef{act2}, the curvature of the
AdS space is:
\begin{equation}
\hL^2=L^2\left(1-\frac{2}{3}\a_3+\cdots\right)\,. \labell{curv1}
\end{equation}
Again, this curvature is dependent on the microscopic details of the
quantum gravity theory.

The key observation, which we review here, is that the two
dimensionless parameters identified above are simply related to
parameters characterizing the dual CFT. First, we recall that the
conformal anomaly of a four-dimensional CFT can be identified by
putting the theory in a curved spacetime and observing
\cite{birrell}
\begin{equation}
\langle T^\mu{}_\mu\rangle_{CFT} =\frac{c}{16\pi^2}
I_4-\frac{a}{16\pi^2} E_4\,. \labell{confag}
\end{equation}
Here $c$ and $a$ are the two central charges of the CFT and $E_4$
and $I_4$ correspond to the four-dimensional Euler density and the
square of the Weyl curvature, respectively. Explicitly,
\begin{equation}
E_4= R_{\mu\nu\rho\lambda}R^{\mu\nu\rho\lambda}-4
R_{\mu\nu}R^{\mu\nu}+R^2 \,,\qquad I_4=
R_{\mu\nu\rho\lambda}R^{\mu\nu\rho\lambda}-2  R_{\mu\nu}R^{\mu\nu}
+\frac 13R^2\,. \labell{ejdef}
\end{equation}
Holographic techniques allow precisely the same expression to be
calculated with the result \cite{holo4,ode,bng},
\begin{equation}
\langle T^\mu{}_\mu\rangle_{holo} =
-\frac{\hL^3}{16\lp^3}
\,\left(E_4-I_4\right)+\frac{\hL\,L^2}{4\lp^3}\,\a_3
\left(E_4+I_4\right)\,, \labell{confa}
\end{equation}
Hence comparing \reef{confag} and \reef{confa}, we arrive at
\begin{equation} \frac{L^3}{\lp^3}
\simeq\frac{c}{\pi^2}\left(1-\frac{3}{8}\frac{c-a}{c}\right)\,,\qquad
\a_3\simeq\frac{1}{8}\frac{c-a}{c}\,. \labell{relate}
\end{equation}
In these expressions, we have used our assumption of a sensible
derivative expansion, which dictates that $\a_3\ll1$.

One conclusion then is that if we require the quantum gravity theory
is described by a low energy action with sensible derivative
expansion, we are restricted to consider CFT's for which
 \beq
 c\sim a\gg1\qquad{\rm and}\qquad |c-a|/c\ll 1\,.
 \labell{restrict}
 \eeq
Further, the effective action is expected to contain further higher
curvature terms and the dimensionless coefficients appearing in
these interactions would be related to new parameters characterizing
the CFT --- for example, see \cite{collide}. Our assumption of a
sensible derivative expansion then restricts the size of these
parameters, \ie the CFT's of interest should have these parameters
being proportional to inverse powers of the central charge $c$.

Above, we observed that the AdS/CFT correspondence dictates the
values of the leading parameters in the effective gravity action
terms of the central charges of the dual CFT according to
\reef{relate}. Hence if the central charges of the CFT are known
(and the inequalities \reef{restrict} are satisfied), we can be
confident of the precise form of this effective action \reef{act2}
to leading order, even if we do not understand the microscopic
details underlying the quantum gravity theory. Then if we are
careful to respect the limitations of the derivative expansion, we
can work reliably with the gravity action \reef{act2} to determine
the properties of the CFT using the standard AdS/CFT correspondence.
Note that only the dimensionless ratios in \reef{relate}, but not
the Planck length $\lp$, appear in any physical results for the CFT.
Of course, this is in accord with the fact that for a supersymmetric
CFT, supersymmetry combines with diffeomorphism and conformal
invariance to completely dictate the form of the two- and
three-point correlators of the stress-energy tensor in terms of
these two central charges, $a$ and $c$ \cite{super1}. Hence while we
can reproduce these correlators with the dual gravity action
\reef{act2}, the latter also allows us to calculate more interesting
properties, such as thermal transport coefficients of the CFT. One
interesting example is the shear viscosity \cite{vi2,kp}
\begin{equation}
\frac \eta s=\frac{1}{4\pi}\left(1-\delta+\cdots\right)\,,
\labell{etas2}
\end{equation}
where we have introduced
\begin{equation}
\delta\equiv \frac{c-a}{c} =8\alpha_3+\cdots\,. \labell{deft}
\end{equation}
Hence the sign of $\delta$ in the CFT or of the $R_{abcd}R^{abcd}$
term in effective gravity action determines whether or not the
viscosity bound \reef{bound} is respected or violated at this order.
In particular, the bound is violated if $c>a$.

Of course, according to the standard AdS/CFT dictionary, the metric
is dual to the stress-energy tensor of the CFT and so with the
gravity action \reef{act2}, we are restricted to study the
properties of this one operator. In general, we should expect the
full CFT will have a spectrum of interesting operators, possibly
including a variety of relevant, irrelevant and marginal operators.
The latter would then be dual to other fields which may also play an
interesting role in the gravity theory. Hence our preceding
conclusions may seem somewhat naive since we have restricted the
discussion to the pure gravity sector of the theory. Therefore we
must show that such operators do not effect our conclusions.

As an example, consider the case where the gravity theory that
contains a number of scalars $\phi^k$. As above, we assume that the
effective gravity theory is described by a sensible derivative
expansion. In principle, a large number of four-derivative terms
could appear in the effective action but as described in appendix
\ref{redef}, field redefinitions can be used to greatly simplify the
action. The final action can be written as
 \beqa
 I&=&\frac{1}{2\lp^3}\int d^5x
\sqrt{-g}\left[\,\calu\left(\phi^m\right) + R
-\calk_{ij}(\phi^m)\,\nabla\phi^i\cdot\nabla\phi^j \right.
 \labell{act3}\\
&&\qquad\left. + \cala_3\left(\phi^m\right)
R_{abcd}R^{abcd}+\calb\left(\phi^m, \nabla_a\phi^m,
\nabla_a\nabla_b\phi^m\right)+\cdots\right]\,.
 \nonumber
 \eeqa
While more details are provided in the appendix, $\calb$ combines
the remaining four-derivative interactions which explicitly contain
derivatives of the scalar fields. An important point is that all of
these interactions contain at least two factors of scalar
derivatives. Then, since we are treating these terms perturbatively
within the derivative expansion and the scalars will be constant in
the leading solutions of interest, they remain constant at the next
order. Hence we may ignore these terms for the remainder of the
discussion.

In describing the rest of the terms in \reef{act3}, we should begin
by saying that we have adopted the convenient (supergravity)
convention where the scalar fields $\phi^i$ are dimensionless.
Below, we argue that the scalars vanish in the AdS$_5$ vacuum and so
we may assume that all of the expressions in the action are
nonsingular at $\phi^i=0$. Hence we can express each of the
coefficient functions in terms of a Taylor series:
 \beqa
\calu\left(\phi^m\right)&=&\frac{12}{L^2} \left( 1+u_i\phi^i +
u_{ij}\phi^i\phi^j + u_{ijk}\phi^i\phi^j\phi^k +\cdots\right)\,,
\labell{pot1}\\
\calk_{ij}\left(\phi^m\right)&=& k_{ij}+k_{ijk}\phi^k +
k_{ijkl}\phi^k\phi^l + k_{ijklm}\phi^k\phi^l\phi^m +\cdots\,,
\labell{kin1}\\
\cala\left(\phi^m\right)&=&L^2 \left(\, \alpha_3+a_i\phi^i +
a_{ij}\phi^i\phi^j + a_{ijk}\phi^i\phi^j\phi^k +\cdots\right)\,.
\labell{pot2}
 \eeqa
Now in keeping with our assumption of the derivative expansion
above, a second key assumption here is that:\\
{\it All of the coupling coefficients in each of \reef{pot1},
\reef{kin1} and \reef{pot2} above are of the same order (with
the exception of $u_i$).}\\
\noindent That is, all of the couplings $u_{ij\cdots}$ in
\reef{pot1} and $k_{ijk\cdots}$ in \reef{kin1} may be of order one
(or higher order in $\lp^2/L^2$), with the exception of $u_i$ --
which we address below. Similarly, $\alpha_3$ and all of the
subsequent couplings $a_{ij\cdots}$ in \reef{pot2} are assumed to be
of order $\lp^2/L^2$ (or higher). Of course, each of these couplings
may in general be a complicated function of $\lp^2/L^2$ and so here
we are demanding that $\alpha_3$ and $a_{ij\cdots}$ do not have
order one (or order $\lp/L$) contributions. Within this framework,
the corresponding scalar masses are of the order of the AdS
curvature scale, \ie $m_k^2\sim 1/L^2$. Hence each of the dual
scalar operators $\calo_k$ has a conformal dimension of order one.
These operators may be relevant, irrelevant or marginal. An exactly
marginal operator is an exceptional case, which will receive
detailed consideration below. As before, the ellipsis in \reef{act3}
corresponds to six- and higher derivative terms which implicitly are
suppressed at least by couplings of order $\lp^4/L^4$, as in the
previous discussion.

The dual theory is a conformal field theory, which again implies
that at any order in the derivative expansion, the gravity theory
\reef{act3} admits an AdS$_5$ vacuum. Further, in the conformal
vacuum, the expectation value of any of the operators must vanish,
\ie
\begin{equation}
\langle \calo_k\rangle_0=0\,,\quad{\rm as\ well\ as}\ \ \ \langle
T_{\mu\nu}\rangle_0=0\,. \labell{vacuum}
\end{equation}
%R3 edits below on vev
This property is reflected in the gravity theory with the vanishing
of the dual scalar fields in the AdS$_5$ vacuum. A possible
exception to this conclusion arises with an exactly marginal
operator. In principle, the corresponding massless scalar in the
dual gravity theory can take on any constant value. However, we will
define this expectation value of the scalar field to be zero for the
vacuum that we are studying here.

Let us now turn to the special case of the exceptional couplings
$u_i$. For the AdS$_5$ space to be a solution with $\phi^i=0$ at
leading order in the derivative expansion, \ie dropping the
curvature-squared and higher order terms, it must be true that
$u_i=0$ at this order. However, when curvature-squared term is
included, the scalar equations of motion yield
\begin{equation}
\left[\frac{\delta\calu}{\delta\phi^i}+\frac{\delta\cala}{\delta\phi^i}
\, R_{abcd}R^{abcd}\right]_{\phi^k=0}=\frac{12}{L^2}\,u_i+L^2\,a_i\,
\frac{40}{L^4}=0 \,. \labell{vacuum2}
\end{equation}
assuming an AdS$_5$ background with vanishing scalars. Hence we find
that consistency demands that the scalar potential contains linear
couplings of order $\lp^2/L^2$:
\begin{equation}
u_i=-\frac{10}{3}\,a_i\, . \labell{linear}
\end{equation}
An alternate interpretation would be that if we set $u_i=0$ then the
AdS$_5$ solution is stable to leading order but in general the
appearance of the curvature-squared term will then cause the
scalars to acquire an expectation value of order $\lp^2/L^2$ in the
vacuum. We are simply redefining the scalars to absorb this constant
shift in our approach.\footnote{ An alternate approach would be to
use the freedom of field redefinitions so that the square of the
Weyl tensor, rather than of the Riemann tensor, appears in the
effective action \reef{act3}. Then because the AdS vacuum has
vanishing Weyl curvature, the scalar equations of motion would be
unaffected by the curvature-squared term and $u_i$ would remain zero
at this order. Note that in this approach, the AdS curvature would
also match precisely the scale $L$ appearing in the action. However,
the additional $R_{ab}R^{ab}$ and $R^2$ interactions, appearing in
the Weyl-curvature squared, would modify the holographic anomaly
\reef{confa} in precisely such a way to reproduce the same
expressions as in \reef{relate}.}

Although the scalars vanish in the AdS$_5$ vacuum, one can expect
that the curvature-squared term will source the various scalar
fields in more general backgrounds. However, one would still have
that $\phi^k\sim a_k\sim \lp^2/L^2$ in such a background. We must
consider two particular examples in our discussion. The first
relevant example would be a black hole background and this effect
implies that in a thermal bath the dual operators acquire
expectation values $\langle\calo_k\rangle \sim a_k$. The second
relevant case comes from the holographic calculation of the
conformal anomaly \reef{confa}. While the precise background is
typically not specified in these calculations, implicitly, one must
be working with more general backgrounds where, in particular, the
Weyl curvature is nonvanishing. Hence we must argue that even though
the scalars may have nontrivial profile at order $\lp^2/L^2$, this
will not affect the holographic calculations of the conformal
anomaly or the thermal behaviour of the CFT. With a careful
consideration below, we will show that the nontrivial scalars can
only modify the results at order $\lp^4/L^4$. Our general argument
was originally formulated in a slightly different context in
\cite{mps}.

As an explicit example, let us consider the calculation of shear
viscosity \cite{calc1,ks,review1}.  A key step would be calculating
the effective quadratic action for the various graviton
fluctuations, \ie the shear, sound and transverse modes, in the
black hole background. The nontrivial scalars can effect these modes
in two ways. First they explicitly appear in the action. However,
contributions of terms quadratic or higher powers in $\phi^i$ would
be suppressed by $\lp^4/L^4$ or higher powers. There are two
possible sets of linear terms in $\calu(\phi^m)$ and in
$\cala(\phi^m)$ but, as discussed above, the couplings for both of
these are already order $\lp^2/L^2$ and so they only contribute with
an overall factor of $\lp^4/L^4$. Secondly, the nontrivial scalars
will modify the background geometry through Einstein's equations,
but similar reasoning shows that the modifications of the metric
would again be order $\lp^4/L^4$. Hence, even though the scalars
themselves appear at order $\lp^2/L^2$, their effect is only felt by
the graviton modes at order $\lp^4/L^4$. Hence the calculation of
$\eta/s$ can be reliably made at order $\lp^2/L^2$ while ignoring
all of the scalar fields, \ie with the effective gravity action
\reef{act2}. The same general argument applies to calculations of
other thermal properties from the black hole background or of the
holographic conformal anomaly.

Next, we make a few comments on the extension of our discussion to
include vectors in the gravity theory --- see also appendix
\ref{redef}. If we consider some number of Abelian gauge fields, the
vectors are dual to conserved currents and the corresponding $U(1)$
gauge symmetries are identified with global symmetries in the CFT
\cite{adscft}. A complete discussion of the contributions of these
gauge fields to the four-derivative gravitational action would be
quite lengthy and equally tedious and so we only remark on salient
points. First, we restrict the discussion to having only constant
gauge fields at leading order in the background. That is, we are
only considering the case of vanishing chemical potentials. Next, it
is relatively easy to show that the majority (\ie all but one) of
the new four-derivative interactions are at least quadratic in the
field strengths of these gauge fields. Hence an argument similar to
that below \reef{act3} applies here as well, with the conclusion
that these terms are irrelevant at this order, as long as we are
considering backgrounds where the vectors are constant. However,
given a set of $U(1)$ gauge fields $A^i_a$, there is one
four-derivative coupling which cannot be dismissed by this argument,
namely,
 \beq
I'=\frac{1}{2\lp^3}\int L^2\,d_i\,A^i\wedge R^a{}_b\wedge R^b{}_a\,.
 \labell{wedgy}
 \eeq
In keeping with the derivative expansion, these terms, which are
linear in the gauge fields, are characterized by a set of
dimensionless constants $d_i\sim\lp^2/L^2$. Note that we require
that under local gauge transformations, $I'$ only produces a surface
term and so even if the gravity theory contains scalars, we cannot
replace the constants $d_i$ by general functions $\cald_i(\phi^m)$.
An interaction of this form plays an interesting role in describing
the anomaly for the $U(1)_R$ current in supersymmetric CFT's
\cite{anom,anom2}. In fact, in the context of $\caln=2$ gauged
supergravity, supersymmetry connects this interaction \reef{wedgy}
to an $R_{abcd}R^{abcd}$ term \cite{sugra,sugra2}. Since this
interaction is linear in the gauge potential, it will induce a
nontrivial profile in a background where $R^a{}_b\wedge R^b{}_a$ is
nonvanishing. However, this combination of curvatures vanishes both
for the AdS$_5$ vacuum and an AdS$_5$ black hole background and so
no profile is induced for these backgrounds. Of course, this result
is in keeping with the intuition that a finite charge density is not
induced by introducing a finite temperature alone. We should also
consider the nontrivial backgrounds implicit in calculating the
holographic conformal anomaly \reef{confa}. In general, we expect
that a nontrivial profile can be induced for the gauge potentials in
this case but we would still only find that $A^k\sim d_k\sim
\lp^2/L^2$ in such a background. Hence following arguments analogous
to those presented to dismiss the effect of nontrivial scalar
profiles, we would again find that the nontrivial gauge potentials
can only modify the anomaly calculation at order $\lp^4/L^4$.
Therefore the calculations of both the thermal properties and of the
conformal anomaly would remain unaffected by the appearance of
additional vector fields. Hence our conclusion once again is that
these calculations can be reliably made at order $\lp^2/L^2$ with
the effective gravity action \reef{act2}, while ignoring any matter
fields in the gravitational theory.\footnote{Of course, the
interesting question of corrections in the presence of a chemical
potential would require a detailed analysis of the higher order
gauge field interactions.}

In closing this section, we return to the special case of an exactly
marginal (scalar) operator. As already mentioned above, the dual
scalar field $\phi^M$ is precisely massless and so it can in
principle be set to any arbitrary value in the AdS$_5$ vacuum. This
property implies special relations between the couplings for
$\phi^M$ in the effective action \reef{act3}, \ie
\begin{equation}
\left.\left(\frac{\delta\ \
}{\delta\phi^M}\right)^n\left[\calu(\phi^k)+\cala(\phi^k)
\,\frac{40}{L^4} \right]\right|_{\phi^k=0} =0 \,, \labell{margins}
\end{equation}
where $40/L^4$ corresponds to $R_{abcd} R^{abcd}$ in AdS$_5$, as in
\reef{vacuum2}. Unfortunately, these relations make the discussion
somewhat more complicated than necessary. So instead, we make a
field redefinition such that the effective action \reef{act3} takes
the form
 \beq
 I=\frac{1}{2\lp^3}\int d^5x
\sqrt{-g}\left[\,\tilde{\calu}\left(\phi^m\right) + R
-\tilde{\calk}_{ij}(\phi^m)\,\nabla\phi^i\cdot\nabla\phi^j  +
\tilde{\cala}_3\left(\phi^m\right) C_{abcd}C^{abcd}+\cdots\right]\,,
 \labell{act35}
 \eeq
where $C_{abcd}$ is the Weyl curvature in five dimensions. Since the
Weyl curvature vanishes in the AdS$_5$ vacuum, the only restriction
is that the scalar potential $\tilde{\calu}\left(\phi^m\right)$ is
completely independent of $\phi^M$. Note, however, that in general
$\tilde{\cala}_3\left(\phi^m\right)$ remains a function of $\phi^M$
without any restrictions. Hence, even though the couplings in
$\tilde{\cala}_3$ are naturally of order $\lp^2/L^2$ as in
\reef{pot2}, this suppression could be overcome if the massless
scalar has a very large expectation value, \ie $\tilde{\cala}_3\sim
\calo(1)$ for large $\phi^M$. More generally, since $\phi^M$ can
become arbitrarily large, it can produce effective coupling
coefficients for the higher derivative terms which are not
suppressed as we initially assumed. Hence, our assumption of a
sensible derivative expansion will implicitly restrict us to a
limited region of the parameter space.
%R3 dropped last clause referring to $\langle\calo_M\rangle$

%R3 edits below -- not moduli space for CFT
Of course, it is generally expected that a CFT with exactly marginal
operators will be an exceptional case and in the absence of exactly
marginal operators, these considerations are not required. However,
this issue naturally arises in many string realizations of the
AdS/CFT correspondence where the dilaton, \ie the string coupling,
is dual to an exactly marginal operator.

In particular, this is the case for the string theory construction
which Kats and Petrov \cite{kp} suggested produces a violation the
KSS viscosity bound \reef{bound}. In this context, the gravitational
theory is Type IIb string theory on a AdS$_5\times S^5/\zet_2$
background, which can be viewed as the decoupling limit of $\nc$
D3-branes overlapping with a coincident collection of four D7-branes
and an O7-brane \cite{fayspa}. The dual CFT is four-dimensional
$\caln=2$ $Sp(\nc)$ super-Yang-Mills coupled to 4 hypermultiplets in
the fundamental representation and 1 hypermultiplet in the
antisymmetric representation. The central charges for this gauge
theory are \cite{at}:
 \beq
c=\frac{N_c^2}{2}+\frac{3\,N_c}{4}-\frac{1}{12}\,, \qquad\quad
a=\frac{N_c^2}{2}+\frac{N_c}{2}-\frac{1}{24}\,.
 \labell{center2}
 \eeq
Now for large (but finite) $\nc$, the central charges satisfy both
of the inequalities in \reef{restrict} and so it seems that we can
confidently apply the results calculated from the five-dimensional
effective action \reef{act2} with the gravitational couplings fixed
by \reef{relate}. Further, as noted by \cite{kp}, $c>a$ and so the
shear viscosity \reef{etas2} is
\begin{equation}
\frac{\eta}{s}=\frac{1}{4\pi}\left(1-\frac{1}{2\nc}+\cdots\right)\,,
\eqlabel{etakp}
\end{equation}
which violates KSS bound \reef{bound}.

However, before accepting this result, we must first consider that
in this construction, the string coupling $\gs$ remains a free
parameter. That is, this corresponds to the case of an exactly
marginal operator which is dual to the dilaton. As usual then, the
results for the CFT can be considered in a double expansion in both
inverse powers of $\nc$ and of the 't Hooft coupling $\lambda$.
Alternatively, we can think that the corrections to effective
gravity action are governed by two independent scales: the Planck
length $\lp$ and the string length $\ls$. Hence, we must make sure
that the higher curvature corrections beyond those explicitly shown
in \reef{act2} are sufficiently suppressed according to the assumed
derivative expansion. As we discuss in section \ref{plasma}, there
will be no curvature cubed interaction. There is a universal term
quartic in curvatures which appears in any (closed) superstring
theory \cite{four}. It is known that this term corrects the ratio of
viscosity-to-entropy-density at $\calo(\lambda^{-3/2})$
\cite{bls,f2,fu1}. Hence as noted in \cite{kp}, in order for the
correction in \eqref{etakp} to dominate, we must have
\begin{equation}
\frac 1\nc\gg \lambda^{-3/2}\qquad \Rightarrow\qquad \lambda\gg
\nc^{2/3}\,. \labell{dom}
\end{equation}
The full correction to $\eta/s$ from the $R^4$ term also contains a
contribution at $\calo( \lambda^{1/2}/\nc^2)$, as well as various
nonperturbative corrections \cite{mps,fu2}. While the latter play no
role in the present discussion, formally requiring the first
correction to be subdominant yields $\lambda^{1/2}\ll\nc$. While the
previous interaction can be associated with the closed string
sector, one might also ask if the calculations could be
significantly effected by $R^4$ interactions induced by the branes.
As explained in appendix \ref{string}, such higher curvature terms
will be subdominant in the derivative expansion. In particular, a
D7-brane induced $R^4$ term would be accompanied by an additional
suppression factor of $\gs V_3 \ls^2/V_5$. In the present case with
$V_3\sim L^3$ and $V_5\sim L^3$, such an $R^4$ interaction would
only contribute corrections at order $1/(\lambda \nc)$. The final
conclusion is that the Kats and Petrov result \reef{etakp}
calculated with a five-dimensional effective action \reef{act2} is
reliable within in a certain parameter regime \reef{dom} and that we
have at least limited violations of the KSS bound \reef{bound} in
string theory.

In the above string theory example, the curvature-squared term can
be associated with the world-volume action of the D7-branes
\cite{at,anom2}. In appendix \ref{string}, we have added a
discussion which provides a schematic understanding of the origin of
this term.

Note the requirement \reef{dom} is compatible with conventional
restrictions implicit in considering the classical gravity limit of
the AdS/CFT correspondence in string theory. That is, we have
$1\ll\lambda\ll\nc$ from requiring $\ls^2/L^2\ll 1$ to minimize
stringy contributions in the derivative expansion and $\gs\ll 1$ to
minimizes string loop contributions. While the derivative expansion,
and hence $\lambda\gg1$, is central to the present effective action
approach, there is no need to give a separate account of loop
contributions. That is, using the five-dimensional effective action
\reef{act2} did not require a detailed understanding of the
underlying microscopic origin of each of the couplings in the full
quantum gravity theory. Rather we advocated that if the CFT central
charges were given, we could use the AdS/CFT dictionary to fix the
gravitational couplings according to \reef{relate}. Of course,
consistency also required that these central charges satisfy the
inequalities given in \reef{restrict}. With this approach, there is
no reason that we could not consider the above or other string
theory constructions where the string coupling is strong, \ie
$\gs\sim1$, which implies that $\lambda\sim\nc$ or $\ls\sim\lp$. In
particular, we can apply this approach to evaluate the thermal
behaviour of CFT's holographically described by the F-theory
constructions of \cite{at}. The case considered by Kats and Petrov
corresponds to one of these constructions and in fact, it is the
only case with a marginal coupling. In the remaining cases, there
are no marginal couplings and the string coupling is pinned at
$\gs\sim1$. Further, as discussed in the following section, with
large (but finite) $\nc$, the central charges again satisfy the
inequalities in \reef{restrict} and so the shear viscosity
\reef{etas2} yields new violations of the KSS bound \reef{bound}
since $c>a$ in each of these cases.

\section{$(c-a)$ in superconformal gauge theories}\label{gauge}

From the results of the previous section, we can conclude that if
the central charges of a four-dimensional superconformal gauge
theory satisfy the two inequalities in \reef{restrict}, then we can
reliably describe the theory with a gravity dual with a
five-dimensional effective action \reef{act2} in which the
gravitational couplings fixed by \reef{relate}. Further, the shear
viscosity is given by \reef{etas2} and the superconformal theory
will violate the KSS bound \reef{bound} provided that
\begin{equation}
\delta=\frac{c-a}{c}>0\,. \labell{dll1}
\end{equation}
Hence in this section, we explore the central charges of
superconformal gauge theories based on simple Lie groups with
various matter fields. We only consider the gauge group $G$ to be a
classical Lie group since we wish to take a large-$N_c$ limit so
that the first inequality in \reef{restrict} will be satisfied. We
discuss two sets of theories, first those in which the gauge
coupling is an exactly marginal operator and secondly models defined
as isolated SCFTs. The gauge theories under consideration have
either $\caln=2$ or $\caln=1$ supersymmetry in four dimensions; some
of them have a known string theory dual while others do not (at this
stage). Quite surprisingly, we find that in all of these models
$\delta\ge 0$, which would seem to indicate a violation of the KSS
bound \reef{bound}. However, generically, $\delta\sim 1$ as
$N_c\to\infty$ and so the second inequality in \reef{restrict} is
not satisfied. Therefore those theories do not have a gravity dual
with a controllable derivative expansion, which is required for
\reef{etas2} to be valid. A similar analysis was carried out by Yuji
Tachikawa and Brian Wecht \cite{same}. Recently, \cite{parn}
presented a complementary analysis of super-QCD with various
relevant superpotentials. Related calculations also appear in
\cite{elias}.
%R3 add ref to elias' paper

A superconformal gauge theory has an anomaly free global $U(1)_R$
symmetry. The central charges, $a$ and $c$, are relatively easy to
determine as they are related to gravitational anomalies in this
global symmetry. Consider a superconformal gauge theory with a gauge
group $G$ and matter multiplets in representations $\{R_i\}$. Let
$r_i$ denote the $R$-charges of the matter chiral
multiplets\footnote{We use $\caln=1$ susy representations to
describe theories with extended supersymmetry as well.} in the
representation $R_i$. It was found in \cite{pc}  that
\begin{equation}
\begin{split}
c-a=&-\frac{1}{16} \left( \dim G +\sum_i \left(\dim R_i\right)
(r_i-1)\right)\,,\\
c=&\frac {1}{32}\left(4\left(\dim G\right)+\sum_i\left(\dim
R_i\right)(1-r_i)\left(5-9(1-r_i)^2\right)\right)\,.
\end{split}
\labell{ca}
\end{equation}
Thus computation of $\delta$ reduces to the identification of the
anomaly-free $U(1)_R$ symmetry of the gauge theory at a
superconformal fixed point. Our approach to this question depends
whether the gauge coupling is marginal or the theory is at an
isolated fixed point.

\subsection{Superconformal gauge theories with exactly marginal gauge coupling}

Let us begin with the identification of the anomaly-free $U(1)_R$
symmetry for the case where the gauge coupling is exactly marginal.
Resolving this question is straightforward in this case as it can be
shown a $U(1)_R$ symmetry with classical assignment of the
$R$-charges is anomaly free, given the vanishing of the one-loop
perturbative $\b$-function. Consider classical assignment of
$R$-charges, \ie all matter superfields have $r_i=\frac 23$ and a
vector superfield has $r_{\adj}=1$. The superconformal algebra then
implies that anomalous dimensions of chiral superfields ($\chi sf$)
must vanish. That is, the  vanishing of the  NSVZ exact perturbative
$\b$-function, which is equivalent for zero anomalous dimensions to
vanishing of one-loop perturbative $\b$-function,
\begin{equation}
0=\b_{\caln=1}(g)\propto \left(\frac 32 \T(\adj)-\frac 12 \sum_{i\in
\chi s f} \T(R_i)\right)\,, \eqlabel{vanbeta}
\end{equation}
guarantees that the classical $R$-charge assignment is in fact
anomaly-free:
\begin{equation}
\begin{split}
\langle\del_\mu j^\mu_R \rangle\ &\propto \biggl(r_{\adj}\
\T(\adj)+\sum_{i\in \chi sf} (r_i-1)\ \T(R_i)\biggr)
=\biggl(\T(\adj)-\frac 13\sum_{i\in \chi sf}  \T(R_i)\biggr)\\
&\propto \b_{\caln=1}(g)=0\,.
\end{split}
\end{equation}
In these expressions, $\T(\adj)$ and $\T(R_i)$ are group indices of
the adjoint representation and a $\chi sf$ representation $R_i$ in
$G$, \eg see \cite{tern} for explicit values. We only consider
non-chiral theories in the following.

\subsubsection{$SU(N_c)$}
Since $\T(\adj)=2N_c$, to satisfy the vanishing of $\b$-function as
$N_c\to \infty$, we can consider (besides adjoint) only fundamental,
symmetric and antisymmetric representations for the $\chi s f$ ---
any other representation has an index growing at least as
$\calo(N_c^2)$ as $N_c\to \infty$.

Suppose we have $n_{adj}$ $\chi sf$ in the adjoint representation,
$n_f$ flavors\footnote{Recall that for a chiral representation, one
flavor is the sum of two conjugate representations.} in the
fundamental representation, $n_{sym}$ flavors in the symmetric
representation and $n_{asym}$ flavors in the anti-symmetric
representation.   Then, the vanishing of the NSVZ $\b$-function
implies
\begin{equation}
0=\frac 32 \cdot 2N_c-\frac 12\biggl(n_{adj}\cdot 2N_c+2n_f\cdot
1+2n_{sym}\cdot (N_c+2)+2n_{asym}\cdot (N_c-2)\biggr)\,,
\labell{sun}
\end{equation}
which we can rearrange to yield
 \beq
n_f=(3-n_{adj}-n_{sym}-n_{asym}) N_c+2(n_{asym}-n_{sym})\,.
 \labell{nfff}
 \eeq
Using this result, we can rewrite $(c-a)$ in \eqref{ca} as
\begin{equation}
c-a=\frac{N_c^2}{16}\left(1-\frac
13(n_{adj}+n_{sym}+n_{asym})\right)
+\frac{N_c}{16}(n_{asym}-n_{sym})+ \frac{1}{16} \left(1-\frac 13
n_{adj} \right)\,. \labell{casu}
\end{equation}
Since $c\sim \calo(N_c^2)$, requiring that $\delta\ll1$ as $N_c\to
\infty$ necessitates
\begin{equation}
n_{adj}+n_{sym}+n_{asym}=3\,, \labell{con1}
\end{equation}
which along with $n_f\ge 0$ further implies that
\begin{equation}
n_{asym}-n_{sym}\ge 0\,. \labell{con2}
\end{equation}

It is easy now to enumerate all the models with $G=SU(N_c)$ and
$\delta\ll 1$ as $N_c\to \infty$, as shown in the following table.
\begin{table}[ht]
\centering
\begin{tabular}{|c|c|c|c|}
    \hline
      & $(n_{adj},n_{asym},n_{sym},n_f)$    &   $c-a$  & $\delta$     \\
    \hline
 (a)&(3,0,0,0) & 0 & 0   \\
    \hline
(b)&(2,1,0,1) & $\frac{3N_c+1}{48}$ & $\frac{1}{4N_c}+\calo(N_c^{-2})$   \\
    \hline
(c)&(1,2,0,2) & $\frac{3N_c+1}{24}$ & $\frac{1}{2N_c}+\calo(N_c^{-2})$   \\
    \hline
(d)&(1,1,1,0) & $\frac{1}{24}$ & $\frac{1}{6N_c^2}+\calo(N_c^{-4})$   \\
    \hline
(e)&(0,3,0,3) & $\frac{3N_c+1}{16}$ & $\frac{3}{4N_c}+\calo(N_c^{-2})$   \\
    \hline
(f)&(0,2,1,1) & $\frac{N_c+1}{16}$ & $\frac{1}{4N_c}+\calo(N_c^{-2})$   \\
    \hline
\end{tabular}
\end{table}
%
%\noindent
Notice that model $(a)$ has a matter content corresponding to
$\caln=4$ susy (and  as a result $\delta_{(a)}=0$). Similarly,
models $(c)$ and $(d)$ have a matter content corresponding to
$\caln=2$ susy. In principle, the five models $(b)$ through $(f)$
are described by a gravity dual with the effective action
\reef{act2}. Further, we note that $\delta>0$ for each of these
models and so they would seem to give violations of the KSS bound
\reef{bound}. However, the gauge coupling is marginal in all of
these models and so we would have to make sure there is a regime in
which $\delta$ gives the dominant correction to the ratio of the
shear-viscosity-to-entropy-density \reef{etas2}. As discussed at the
end of section \ref{dual}, if we imagine that the gravity dual comes
from a string theory construction, this should be possible for
models $(b,c,e,f)$ with $\delta\sim1/\nc$ if the inequality
\reef{dom} is satisfied. However, for model $(d)$ with
$\delta\sim1/\nc^2$, we should note that the $R^4$ interactions are
also expected to contribute (positive) corrections at
$\calo(\lambda^{1/2}/\nc^2)$. The latter would always dominate since
$\lambda\gg1$ is also required for a sensible derivative expansion.
However, the four superconformal gauge theories $(b,c,e,f)$
potentially have string theory duals which would produce violations
of the KSS bound.

\subsubsection{$SO(2N_c+1)$ and $SO(2N_c)$}

The analysis proceeds precisely as before. As well as $n_{adj}$
$\chi sf$ in the adjoint (or antisymmetric) representation, we
consider $n_v$ $\chi sf$ in the vector representation and $n_{sym}$
$\chi sf$ in the symmetric representation. For general
$(n_{adj},n_{sym},n_v)$, subject to vanishing $\b$-function, we find
as $N_c\to\infty$
\begin{equation}
\delta=\frac 12\
\frac{3-n_{adj}-n_{sym}}{9-n_{adj}-n_{sym}}+\calo(N_c^{-1})\,,
\labell{so1}
\end{equation}
which is supplemented with the condition
\begin{equation}
n_v\ge 0\,,\qquad \Longrightarrow\qquad 3-n_{adj}-n_{sym}\ge 0\,.
\labell{so2}
\end{equation}
This suggests that while $\delta\ge 0$, the only model with a
controllable gravitational dual is the one with $\delta=0$ and the
$\caln=4$ susy matter content since when $n_{adj}+n_{sym}=3$, the
condition $n_v\ge 0$ also requires that $n_{sym}=0$.

\subsubsection{$Sp(N_c)$}

The analysis is the same as before. Besides $n_{adj}$ $\chi sf$ in
the adjoint (symmetric) representation, we consider $n_f$ $\chi sf$
in the fundamental representation and $n_{asym}$ $\chi sf$ in the
antisymmetric representation.

It is straightforward to establish that $\delta\ge 0$ always as
$N_c\to \infty$ and  to enumerate all the models with $\delta\ll 1$:
\begin{table}[ht]
\centering
\begin{tabular}{|c|c|c|c|}
    \hline
      & $(n_{adj},n_{asym},n_f)$    &   $c-a$  & $\delta$     \\
    \hline
 (a)&(3,0,0) & 0 & 0   \\
    \hline
(b)&(2,1,4) & $\frac{6N_c-1}{48}$ & $\frac{1}{4N_c}+\calo(N_c^{-2})$   \\
    \hline
(c)&(1,2,8) & $\frac{6N_c-1}{24}$ & $\frac{1}{2N_c}+\calo(N_c^{-2})$   \\
    \hline
(d)&(0,3,12) & $\frac{6N_c-1}{16}$ & $\frac{3}{4N_c}+\calo(N_c^{-2})$   \\
    \hline
\end{tabular}
\end{table}

\noindent Notice that model $(a)$ has a matter content corresponding
to $\caln=4$ susy (and  as a result $\delta_{(a)}=0$). Model $(c)$
is that originally identified by Kats and Petrov \cite{kp} and has a
matter content corresponding to $\caln=2$ susy. Models $(b)$ and
$(d)$ provide interesting new candidates for a controllable gravity
dual which again yield violations of the KSS bound \reef{bound}.

\subsection{Isolated superconformal fixed points}

There are several ways to engineer an isolated superconformal fixed
point. In a purely field theoretical construction, we can define an
asymptotically free gauge theory in the UV, which flows to a
strongly coupled interactive conformal fixed point in the IR
\cite{sei1,sei2}. These models have $\caln=1$ supersymmetry.
Alternatively, one can engineer isolated superconformal fixed points
arising from the large number of D3-branes at singularities in
F-theory \cite{ft1,ft2,ft3,ft4,ft5}. The latter have $\caln=2$
supersymmetry. All these theories have non-classical assignment of
$R$-charges of the anomaly-free global $U(1)_R$ symmetry for matter
fields, which implies $\calo(1)$ anomalous dimensions of chiral
superfields --- of course, this is simply a reflection of the strong
coupling at the isolated superconformal fixed point. Unlike the
examples of superconformal fixed points with exactly marginal
coupling discussed above, in the models which we review here with
$\caln=1$ supersymmetry, $\delta$ is always positive but it is not
suppressed by inverse powers of $\nc$ in the large $\nc$ limit.
Therefore these theories will not have a controllable gravity dual
and cannot be proven with the approach considered here to give
counterexamples to the KSS bound.\footnote{Besides models listed
below, we have considered Kutasov-Schwimmer model \cite{ku} and
there again we find $\delta>0$ but $\delta\sim 1$.} On the other
hand, the models of \cite{at} engineered directly in string theory
can violate the KSS bound \reef{bound}.

\subsubsection{Conformal window for $\caln=1$ $SU(N_c)$ gauge theory}

Consider $\caln=1$ $SU(N_c)$ gauge theory with $n_f$ flavors in the
fundamental representation. As shown in \cite{sei1}, for $\frac 32
N_c<n_f< 3 N_c$ the theory flows to a nontrivial superconformal
fixed point in the IR. The matter fields global anomaly-free
$U(1)_R$ charge assignment is as follows \cite{sei1}:
\begin{equation}
r_i=1-\frac{N_c}{n_f}\,, \labell{sus}
\end{equation}
which from \eqref{ca} implies
\begin{equation}
c-a=\frac{N_c^2+1}{16}\,,\qquad \delta=\frac{n_f^2}{7 n_f^2-9
N_c^2}+\calo\left(N_c^{-2},n_f^{-2}\right)\,. \labell{sus1}
\end{equation}
The work of \cite{parn} expands on these results by adding adjoint
matter fields and studying the effect of various superpotential
terms. In these theories, they again find that $\delta$ is always
positive but also order one in the limit of large $\nc$.

\subsubsection{Conformal window for $\caln=1$ $SO(N_c)$ gauge theory}

Consider $\caln=1$ $SO(N_c)$ gauge theory with $n_f$ flavors in the
vector representation. As shown in \cite{sei1}, for $\frac 32
(N_c-2)<n_f< 3 (N_c-2)$ the theory flows to a nontrivial
superconformal fixed point in the IR. The matter fields global
anomaly-free $U(1)_R$ charge assignment is as follows \cite{sei1}:
\begin{equation}
r_i=1-\frac{N_c-2}{n_f}\,, \labell{s0s}
\end{equation}
which from \eqref{ca} implies
\begin{equation}
c-a=\frac{N_c(N_c-3)}{32}\,,\qquad \delta=\frac{n_f^2}{7 n_f^2-9
N_c^2}+\calo\left(N_c^{-1}\right)\,. \labell{s0s1}
\end{equation}

\subsubsection{Conformal window for $\caln=1$ $Sp(N_c)$ gauge theory}

Consider $\caln=1$ $Sp(N_c)$ gauge theory with $2n_f$ flavors in
the fundamental representation. As shown in \cite{sei2}, for $\frac
32 (N_c+1)<n_f< 3 (N_c+1)$ the theory flows to a nontrivial
superconformal fixed point in the IR. The matter fields global
anomaly-free $U(1)_R$ charge assignment is as follows \cite{sei2}:
\begin{equation}
r_i=1-\frac{N_c+1}{n_f}\,, \labell{sps}
\end{equation}
which from \eqref{ca} implies
\begin{equation}
c-a=\frac{N_c(2N_c+3)}{16}\,,\qquad \delta=\frac{n_f^2}{7 n_f^2-9
N_c^2}+\calo\left(N_c^{-1}\right)\,. \labell{sps1}
\end{equation}

\subsubsection{$\caln=2$ superconformal fixed points from F-theory}

Models constructed as $\nc$ D3-branes probing an F-theory
singularity generated by a collection of $n_7$ coincident $(p,q)$
7-branes and resulting in a constant dilaton were classified in
\cite{at}. Classifying the F-theory singularity with the symmetry
group $\calg$, one finds \cite{at}
\begin{table}[ht]
\centering
\begin{tabular}{|c|ccccccc|}
    \hline
$\calg$& $H_0$& $H_1$& $H_2$& $D_4$& $E_6$& $E_7$& $E_8$\\
    \hline
$n_7$& 2& 3& 4& 6& 8& 9& 10\\
%$\gamma$& 6/5& 4/3& 3/2& 2& 3& 4& 6\\
    \hline
\end{tabular}
\end{table}

\noindent We emphasize that the F-theory analysis fully accounts for
the back-reaction of the $n_7$ 7-branes, which generate a deficit
angle $\pi\, n_7/6$ in the internal geometry. Central charges of the
dual four-dimensional $\caln=2$ superconformal gauge theories were
also computed \cite{at} and as $\nc\to\infty$, one has
\begin{equation}
c-a=\frac 14 \left(\frac{n_7}{12-n_7}\right) \nc
-\frac{1}{24}\,,\qquad
\delta=\frac{n_7}{12}\,\frac{1}{\nc}+\calo(\nc^{-2})\,. \labell{caf}
\end{equation}
Notice then that with large but finite $\nc$, each of the models
tabulated above yields $\delta>0$ and $\delta\ll 1$ . Hence they all
have a controllable gravity dual and \reef{etas2} yields a violation
of the KSS bound \reef{bound}. The string coupling remains arbitrary
in the $D_4$ model and so this actually corresponds to a
superconformal gauge theory with an exactly marginal operator. Of
course, this is precisely the case that was examined by Kats and
Petrov \cite{kp}.\footnote{In the present F-theory description, the
O7-plane is resolved as a combination of a (-1,-1) and a (1,-3)
7-brane.}

\section{The strongly coupled quark-gluon plasma?}\label{plasma}

Our analysis in section \ref{dual} demonstrates that the thermal
properties of a large class of conformal gauge theories can be
derived from a simple holographic framework. Of course, one is
tempted to consider how these results might be applied to understand
the strongly coupled quark-gluon plasma, which is currently under
study with experiments at RHIC and soon at the LHC. In this
direction, we would like to generalize a phenomenological approach
originally advocated in \cite{fu2}. The essential first step is to
assume that the QCD plasma is described by an effective conformal
field theory. Given this assumption, this effective CFT will have a
holographic dual according to the AdS/CFT correspondence and if
nature is gracious, the dual theory may be one for which we
calculate. That is, the holographic dual may be approximated by the
five-dimensional Einstein gravity coupled to a negative cosmological
constant, with controllable higher curvature corrections.

In this case, we can ask if the sQGP is described by an effective
CFT within the class of theories whose dual is governed the low
energy action \reef{act2}. We can then treat the parameters
characterizing the CFT, \ie the central charges $a$ and $c$, or
equivalently the dual gravitational parameters $\lp/L$ and
$\alpha_3$, as phenomenological. That is, we can calculate the
properties of the gauge theory plasma from the gravity dual and then
compare the results to experimental observations of QCD to fix the
effective parameters. One interesting property for such a comparison
would be $\eta/s$, as given in \reef{etas2}. As discussed in
\cite{fu2}, if we denote the energy density of the conformal plasma
and of the corresponding free theory as $\vareps$ and $\vareps_0$,
then the ratio $\vareps/\vareps_0$ provides another interesting
quantity for comparison, as the ratio can also be determined by
lattice QCD calculations. Hence our next step is to determine
$\vareps/\vareps_0$ holographically with the effective action
\reef{act2}.

Working to first order in $\alpha_3$ or $\lp^2/L^2$, the equilibrium
state of CFT plasma is encoded in the AdS$_5$-Schwarzschild
background geometry \cite{kp}
\begin{equation}
ds_{T}^2=\frac{r^2}{L^2}\left(-f(r)dt^2+
d\vec{x}^2\right)+\frac{L^2}{r^2}\,\frac{dr^2}{f(r)}\,, \labell{Ts}
\end{equation}
where
 \beq
f(r)=1-\frac{r_0^4}{r^4}+\frac{2}{3}\a_3 +2\a_3\frac{r_0^8}{r^8}\,.
 \label{func1}
 \eeq
The horizon appears at
 \beq
r_H=r_0\,\left(1-\frac{2}{3}\,\a_3\right)\,,
 \label{horizon}
 \eeq
and the plasma temperature corresponds to the Hawking temperature
which is given by
%R3 fixed sign here and factors in the following equations
 \beq
T=\frac{r_0}{\pi L^2}\left(1 -\frac{7}{3}\,  \a_3 \right)\,.
 \labell{temperature}
 \eeq
Next we evaluate the black hole entropy for the solution \reef{Ts}
following the standard approach of \cite{entropy} for gravity
actions with higher curvature corrections. The general expression
takes the form
 \beq
S=-2\pi \oint \frac{\delta\,\call}{\delta
R_{abcd}}\,\hat{\varepsilon}_{ab}\hat{\varepsilon}_{cd}\,\bar{\varepsilon}\,.
\labell{wald}
 \eeq
Of course, in the present case with a planar horizon, the horizon
area diverges and so we calculate the entropy density. Dividing by
the coordinate volume, the final result can be expressed as
 \beq
s=\frac{S}{V_\mt{CFT}}=2\pi\frac{L^3}{\lp^3}\left(\frac{r_H}{L^2}\right)^3
\left[1 + 4L^2\,\a_3\, R^{tr}{}_{tr}\right]_{r=r_H}=
2\pi\frac{L^3}{\lp^3}\left(\frac{r_H}{L^2}\right)^3 \left(1 +
8\,\a_3\right)\,.
 \labell{density1}
 \eeq
Note here that since the curvature above is multiplied by $\a_3$, we
can evaluate it on the leading order solution, \ie $\left.
R^{tr}{}_{tr} \right|_{r=r_0}=2/L^2$. To express this result in
terms of CFT parameters, we use the relations \reef{relate} as well
as our expressions above for the horizon radius \reef{horizon} and
temperature \reef{temperature}. Combining all of these, we arrive at
the final result
 \beq
s\simeq 2\pi^2 c\, T^3\left(1+\frac{5}{4}\frac{c-a}{c}\right)\,.
 \labell{density2}
 \eeq

We would like to compare this result for the entropy density which
is implicitly calculated for strong coupling to the entropy density
of the free field limit. To produce a quantitative result, it turns
out that we must assume that the underlying CFT is supersymmetric.
We begin by noting that the central charges may be written as
\cite{birrell}:
 \beqa
a&=&\frac{124\,N_1+11\,N_{1/2}+2\,N_0}{720}\,,\labell{central2}\\
c&=&\frac{12\,N_1+3\,N_{1/2}+N_0}{120}\,,\nonumber
 \eeqa
where $N_1$, $N_{1/2}$ and $N_0$ denote the number of vectors,
(chiral) fermions and scalars, respectively. While these expressions
assume that these are all massless free fields, the results are
protected in a supersymmetric theory and so also apply at finite
coupling in that case. In a supersymmetric theory, we have an equal
number of bosonic and fermionic degrees of freedom, which we denote
as $N=2N_1+N_0=2N_{1/2}$, and therefore the entropy density is
naturally proportional to $N$. Hence we find the linear
combination\footnote{The same result follows from adding together
the two expressions in \reef{ca} and evaluating the result with all
$r_i=2/3$.}
 \beq
\left(2c-a\right)=\frac{2(2N_1+N_0)+5N_{1/2}}{144}=\frac{1}{32}N\,.
 \labell{linear0}
 \eeq
Now assuming we have a collection of free fields, the entropy
density is easily calculated to be \cite{classic}
 \beq
s_0=\frac{\pi^2}{12}NT^3=\frac{8\pi^2}{3}c\,T^3
\left(1+\frac{c-a}{c}\right)\,.
 \labell{free}
 \eeq
Hence comparing with \reef{density2}, the ratio becomes
 \beq
\frac{s}{s_0}=\frac{3}{4} \left(1+\frac{1}{4}\frac{c-a}{c}\right)\,.
 \labell{ratio2}
 \eeq
Note that we recover the celebrated result $s/s_0=3/4$ with $c=a$
\cite{classic}, in which case the gravity dual reduces to Einstein
gravity (coupled to a negative cosmological constant). Further,
however, the sign of the correction to the ratio here is the
opposite to that for the ratio of shear viscosity to entropy density
\reef{etas2}. Of course for a conformal (or free) field theory, the
energy density and entropy density are simply related as
$\varepsilon=\frac{4}{3}sT$. Hence the result in \reef{ratio2}
applies equally well for the ratio of the energy densities of the
strongly coupled and free theories.

Collecting our results then, all of the CFT's for which \reef{act2}
represents the gravity dual will have the following:
 \beq
\frac{\vareps}{\vareps_0}=\frac{3}{4}\left(1 +
%RM 3/4 --> 1/4
\frac{1}{4}\,\delta\right) \qquad{\rm and}\qquad
\frac{\eta}{s}=\frac{1}{4\pi}\left(1-\delta \right) \,,
 \label{new1}
 \eeq
 \beq
{\rm where}\qquad \delta\equiv\frac{c-a}{c}\,.
 \label{new2}
 \eeq
In principle, the CFT's in the class of interest here have two
independent parameters, $a$ and $c$, but above we have chosen two
quantities which only depend on the combination $\delta$. Thus we
may treat $\delta$ as a phenomenological parameter under the
assumption that the effective CFT describing the QCD plasma lies
within this class. This assumption is then put to the test if both
$\vareps/\vareps_0$ and $\eta/s$ can be constrained by observation
and consistently fit with the same value of $\delta$. We can begin
by using lattice QCD results to fix $\delta$ with the energy
density. Recent studies seem to indicate that energy density should
be in the range $\vareps/\vareps_0\approx0.85\,-\,0.90$
\cite{lattice}. In this case, \reef{new1} yields
%R3 following numbers changed alot and hence words change too
$\delta\approx0.53\,-0.80$ and hence
 \beq
\left.\frac{\eta}{s}\right|_\mt{QCD}\approx0.016\,-\,0.037\ .
 \label{fantasy}
 \eeq
These `corrected' values for $\eta/s$ are significantly lower than
the leading result, \ie the conjectured KSS bound
$\left.\eta/s\right|_\mt{KSS} =1/4\pi \simeq0.08$ \cite{kss}.
%RM new
Even though the `correction' to $\vareps/\vareps_0$ is small, our
fit produced a range of large values for the parameter $\delta$. In
fact, these values are all too large since consistency of the
effective CFT demands that $|\delta|<.5$ \cite{collide}. Hence we
can conclude that the class of holographic models considered here
cannot describe an effective CFT for the QCD plasma and we must
broaden the universality class under consideration.
%RM old:
%However, these results \reef{fantasy} are certainly consistent with
%values emerging from the analysis of RHIC data \cite{rick}. We must
%also note that the value of $\delta$ determined here is not small
%and so higher order corrections at order $\delta^2$ could easily
%modify these results by 50\%.

When considering higher order corrections, it is natural to take into
account higher curvature terms in the effective gravity theory
beyond the curvature-squared term appearing in \reef{act2}.
Naturally the next term to consider would involve a contraction of
three Riemann tensors. The corresponding coupling constant would be
dual to a new CFT parameter in the three point function of the
stress tensor. However, one can argue that this parameter, and hence
the dual gravitational coupling, vanishes for any supersymmetric CFT
\cite{collide}. Since supersymmetry was an underlying assumption in
the analysis above, it is natural then to set the $R^3$ term to
zero.

As already discussed in section \ref{dual}, string theory provides a
specific interaction quartic in curvatures \cite{four}. As
considered there, in situations where the string coupling is a free
parameter, this term does not necessarily enter the action
suppressed by $\lp^6/L^6$. Hence the contributions of this $R^4$
term can be enhanced in certain regimes of the parameter space. For
example in $\caln=4$ super-Yang-Mills, the correction to $\eta/s$ is
$15\zeta(3)/\lambda^{3/2}$ \cite{fu1} and if we evaluate this
contribution with $\lambda=6\pi$, as might be applicable for the QCD
plasma, it could easily compete with $1/\nc$ contributions (with
$\nc=3$ as in QCD). With this observation, we argue that it is not
unreasonable to include the corrections from both the $R^2$ and
$R^4$ terms as making independent and comparable contributions to
the CFT properties, \ie
 \beq
\frac{\vareps}{\vareps_0}=\frac{3}{4}\left(1 +
\frac{1}{8}\Delta+\frac{1}{4}\delta\right) \qquad{\rm and}\qquad
\frac{\eta}{s}=\frac{1}{4\pi}\left(1+\Delta -\delta\right) \,,
 \label{new5}
 \eeq
where $\Delta$ encodes the $R^4$ corrections \cite{fu2}. Within the
context of the phenomenological program advocated above, we have
expanded the class of CFT's which might describe the sQGP and so now
have greater freedom in fitting the observed values of these
quantities. In order to arrive at a constrained or predictive
system, we have to calculate more physical properties of the QCD
plasma. This does not present a real obstacle for the holographic
framework since the effective gravity action allows us to calculate
the corrections for any properties having to do with the
stress-energy tensor. So in particular, we can calculate corrections
to the higher order transport coefficients \cite{higher} and with
these we may be able to produce a constrained set of observables.

\section{Discussion}\label{end}

We examined the effective low-energy description of the
gauge/gravity duality, relevant for discussing thermal and
hydrodynamic properties of strongly coupled conformal gauge theory
plasmas. We argued that as long as the central charges of the CFT
satisfy
 \beq
 c\sim a\gg1\qquad{\rm and}\qquad |c-a|/c\ll 1\,,
 \labell{restrict2}
 \eeq
the dual gravity description should be described with Einstein
gravity coupled to a negative cosmological constant with
perturbative corrections coming from a curvature-squared
interaction. The standard results for the holographic conformal
anomaly \cite{holo4,ode,bng} precisely fix the relevant
gravitational couplings in terms of the central charges. Our
arguments assumed the validity of the effective field theory
description in gravity dual, \ie a reasonable derivative expansion
and generic couplings for any matter fields. These assumptions may
only be satisfied in a particular regime in theories with exactly
marginal operators. In appendix \ref{string}, we use type IIb string
theory, more specifically type IIb supergravity plus probe
D$p$-branes (including leading $\ls^2$ corrections) to establish
that under certain conditions, holographic dualities can indeed be
cast in the framework of the proposed low-energy description.

A primary motivation of our work was to examine the claim by Kats
and Petrov \cite{kp} that the KSS bound \reef{bound} is violated in
a certain string theory model. Our detailed analysis agrees that
their calculations are in fact reliable and the bound is violated in
the regime where $\lambda\gg\nc^{2/3}$, as they already noted. It is
interesting that this restriction establishes the CFT coupling
cannot be arbitrarily small if the KSS bound is to be violated. This
is in keeping with the intuition that bound must not be violated at
weak coupling because the viscosity grows arbitrarily large in the
perturbative regime. This restriction can also be translated into a
limit on how small the string coupling can be if the bound is
%R3 check power
violated in this string theory model, \ie $\gs\gg\nc^{-1/3}$.

Often one also restricts the string coupling $\gs\ll 1$ to be in a
perturbative regime where the microscopic details of the duality can
be well understood, \eg in our schematic discussion in appendix
\ref{string}. However, one important observation in section
\ref{dual} is that these microscopic details are inessential to the
low energy gravity action \reef{act2}. Rather we can use the central
charges to precisely fix the gravitational couplings with
\reef{relate}. Hence, as long as we can evaluate the central charges
from the CFT and they satisfy the inequalities \reef{restrict2}, we
can reliably calculate the leading order corrections in $\delta$
with the effective action \reef{act2}, irrespective of the string
coupling. Hence the result \reef{etas2} for $\eta/s$ is still
dependable for the F-theory models of \cite{at} where the string
coupling is fixed to be order one. As discussed in section
\ref{gauge}, in the limit of large $\nc$, these models provide new
examples where the KSS bound is violated.

In section \ref{gauge}, we also found various new superconformal
gauge theories with $0<\delta\sim 1/\nc$  (as well as $c\gg1$) in
the limit of large $\nc$. Even though no string theory model has
(yet) been constructed which is dual to these gauge theories, by the
arguments of section \ref{dual}, these CFT's will have a
controllable gravity dual described by the action \reef{act2}. Hence
we can be confident that they also represent new examples where the
KSS bound is violated. A caveat in these cases is that the gauge
coupling is precisely marginal and so we expect that the bound will
only be violated in the regime of large 't Hooft coupling. One's
experience with the universal contributions of the $R^4$ interaction
arising in string theory \cite{fu2} suggests that we must require
%R3 check power
$\lambda\gg\nc^{2/3}$, as discussed for the example of \cite{kp}. In
section \ref{gauge}, we also found one example where
$0<\delta\sim1/\nc^2$ but we argued that the theory respects the KSS
bound (in the large $\nc$ limit) since $\lambda$ cannot be tuned to
a regime of where the $R^2$ contribution dominates.

A general feature that we found for all of the superconformal gauge
theories analyzed in section \ref{gauge} was that $\delta$ is
positive. While we focussed there on cases with $\delta\ll1$, all of
our examples of $\caln=1$ theories which flowed to a nontrivial
superconformal fixed point had $\delta$ was positive but
$\delta\sim1$ for $\nc$ large --- the same result applies for the
examples in \cite{parn}. This feature is also the generic behaviour
of the superconformal theories with an exactly marginal gauge
coupling. For example, if we do not insist that $|\delta|\ll1$, then
we see that \reef{casu} still always yields $c-a>0$ for large $\nc$
because of the constraint that $N_f>0$ combined with \reef{nfff}.
However, this generic case yields $c-a\sim \nc^2$ and so
$\delta\sim1$. Although the gravity dual for such a CFT may be
weakly curved, it would seem not to have a controlled derivative
expansion. Therefore while we have found that $\delta>0$ is
generically positive for superconformal gauge theories, the
implications of this observation remain unclear. Further, we should
add that $\delta<0$ can be achieved with a theory of free vector
multiplets with $\caln=0,1,2$ supersymmetry \cite{collide,ya}. Of
course, since these examples are free theories, they will not have a
weakly curved gravity dual.

Our present discussion is limited to considering $|\delta|\ll1$ and
so any of our counter-examples to the KSS bound only produce small
violations of the bound. However, this was simply a technical
limitation arising since we need $|\delta|\ll1$ to reliably
formulate the gravity dual using the techniques of effective field
theory. One might imagine that violations of the KSS bound still
arise when $\delta\sim1$, which as described above is the generic
case, and further that these violations may become arbitrarily large
in this case. However, on general grounds \cite{kss1}, one expects
that $\eta/s$ must remain finite and order one (in units where
$\hbar=1=c=k_\mt{B}$). Hence it is interesting then that basic
considerations of three-point functions in any four-dimensional
supersymmetric CFT seem to restrict $\delta\le1/2$ \cite{collide}.
Further precisely the same bound was found by demanding causality in
a holographic framework where the gravity dual incorporated the
Gauss-Bonnet term as the curvature-squared interaction \cite{vi2a}.
Taken at face value, the latter calculations suggest that the
violations of the KSS bound are limited with $\eta/s\ge 16/100\pi$
for the superconformal gauge theories. However, firmly establishing
a clear lower bound for $\eta/s$ remains an open question.

In section \ref{plasma}, we advocated a phenomenological approach to
applying the AdS/CFT correspondence to understanding the strongly
coupled quark-gluon plasma of QCD. Assuming the sQGP is described by
an effective conformal field theory the latter should be
characterized by a few parameters controlling the aggregate
properties of the plasma. With the AdS/CFT correspondence, these
parameters would then fix the couplings of the dual gravity theory,
\eg as in \reef{relate}. These parameters could then be fixed by
comparing the results determined by holographic calculations with
those emerging from analysis of experimental data, as well as
lattice calculations. By taking into account sufficiently many
quantities, the comparison is constrained and one can concretely
test the assumption that the QCD plasma is described by a CFT within
the universality class defined by a certain family of gravity duals.
With the gravity dual, we are restricting our attention to the
properties of the CFT probed by the stress tensor, however, the
holographic framework allows us to calculate any quantities
originating with this operator. Hence, in principle, there is no
problem in expanding the calculations to a sufficiently broad set of
quantities so that the suggested comparison becomes constrained. At
present, the obstruction to this phenomenological program is that
the experimental data does not yet yield precision results for most
quantities of interest.

Implicit in this discussion is also the assumption that the
effective CFT describing the sQGP is close to Einstein gravity. That
is, the gravity dual is Einstein gravity coupled to a negative
cosmological constant with perturbative corrections coming from a
limited number of higher curvature interactions. Again, this is
simply a technical issue as our present understanding limits our
holographic calculations to producing reliable results within this
framework. The primary motivation to believe that nature could be so
kind as to respect these limitations was that the value for the
shear viscosity emerging from the RHIC data \cite{rick} is unusually
small and even seems to be roughly $1/4\pi$, the universal result
for Einstein gravity duals
\cite{kss1,u1,bls,mps,u3,u4,u5,u55,hong,buchbinder}. The present
discussion may call this motivation into question. Above we found
that the value of $\eta/s$ can become smaller than $1/4\pi$ but
suggested that it will not become too much smaller even if we go
well beyond the regime where corrections to Einstein gravity
can be treated perturbatively. Further, having realized that the KSS
bound can be violated, we observe that if \reef{new5} is
representative then $\eta/s=1/4\pi$ only defines a codimension-one
surface in the space of possible CFT's and so even if this precise
value is found for the sQGP, it is not clear how close the effective
CFT will be to having an Einstein gravity dual.

\section*{Acknowledgments}
We would like to thank Jaume Gomis, Keshav Dasgupta, Ulrich Heinz,
%R3 add kiritsis
Yevgeny Kats, Elias Kiritsis, Jim Liu, Hong Liu, Juan Maldacena,
Steve Shenker and Yuji Tachikawa for interesting discussions.
Research at Perimeter Institute is supported by the Government of
Canada through Industry Canada and by the Province of Ontario
through the Ministry of Research \& Innovation. AB gratefully
acknowledges further support by an NSERC Discovery grant and support
through the Early Researcher Award program by the Province of
Ontario. RCM also acknowledges support from an NSERC Discovery grant
and funding from the Canadian Institute for Advanced Research.

\appendix

\section{Comments on Field Redefinitions}\label{redef}

As remarked in section \ref{dual}, in general, the full CFT will
have a spectrum of interesting operators, each of which will be dual
to an independent field in the gravity theory. These fields will
appear in interactions at all orders in the derivative expansion and
it is interesting to examine how field redefinitions can modify the
higher derivative terms for such fields. For simplicity, we begin
our discussion here by adding a single scalar field to the gravity
dual. However, we will comment on the case of multiple scalars and
other generalizations below. The most general four-derivative action
for gravity coupled to a scalar field $\phi$ (as well as a negative
cosmological constant) is:
 \beqa
 I&=&\frac{1}{2\lp^3}\int d^5x
\sqrt{-g}\left[\,\calu\left(\phi\right) + R
-\calk(\phi)\,\nabla\phi\cdot\nabla\phi \right.
 \labell{act11}\\
&&\quad+ \cala_1\left(\phi\right) R^2+ \cala_2\left(\phi\right)
R_{ab}R^{ab} +\cala_3\left(\phi\right) R_{abcd}R^{abcd}
 \nonumber\\
&&\quad+\calb_{1}\left(\phi\right)\,\nabla\phi\cdot\nabla\phi \,R+
\calb_{2}\left(\phi\right)\,\jsquare\phi\, R
+\calb_{3}\left(\phi\right) \nabla^a\phi\nabla^b\phi \,R_{ab}
+\calb_{4}\left(\phi\right) \nabla^a\nabla^b\phi\,R_{ab}
 \nonumber\\
&&\quad+\calc_{1}\left(\phi\right) (\nabla\phi\cdot\nabla\phi)^2 +
\calc_{2}\left(\phi\right)\,(\jsquare\phi)^2 +
\calc_{3}\left(\phi\right) \nabla\phi\cdot\nabla\phi\, \jsquare\phi
+\calc_{4}\left(\phi\right) \jsquare{}^2\phi
 \nonumber\\
&&\left. \quad +\calc_{5}\left(\phi\right)
\nabla_a\phi\,\nabla^a\jsquare\phi +\calc_{6}\left(\phi\right)
\nabla_a\nabla_b\phi\,\nabla^a\nabla^b\phi
+\calc_{7}\left(\phi\right)
\nabla_a\nabla_b\phi\,\nabla^a\phi\nabla^b\phi\right]\,.
 \nonumber
 \eeqa
In general, one might have expected an additional function
$\calv(\phi)$ to be multiplying the Einstein term, but implicitly we
have eliminated such a coupling with a conformal transformation:
$g_{ab}\rightarrow \calv(\phi)^{-2/3}g_{ab}$. As in section
\ref{dual}, we have adopted the convention that $\phi$ has zero
engineering dimension and we are also assuming that the various
coefficient functions, \eg $\cala_i$, $\calb_i$ and $\calc_i$, are
nonsingular at $\phi=0$. Many of the four-derivative terms above can
be eliminated by simply integrating by parts. For example,
 \beqa
&&\int d^5x \sqrt{-g}\, \calc_{7}\left(\phi\right)
\nabla_a\nabla_b\phi\,\nabla^a\phi\nabla^b\phi
 \labell{examp1}\\
&&\qquad=-\frac{1}{2}\int d^5x \sqrt{-g}
\left(\calc'_{7}\left(\phi\right) (\nabla\phi\cdot\nabla\phi)^2 +
\calc_{7}\left(\phi\right)
\nabla\phi\cdot\nabla\phi\,\jsquare\phi\right)\,,
 \nonumber
 \eeqa
where $\calc'_7\equiv \delta \calc_7/\delta\phi$. In this way, one
can eliminate $\calb_4$, $\calc_4$, $\calc_5$, $\calc_6$ and
$\calc_7$. Hence the general four-derivative action can be reduced
to
 \beqa
 I&=&\frac{1}{2\lp^3}\int d^5x
\sqrt{-g}\left[\,\calu\left(\phi\right) + R
-\calk(\phi)\,\nabla\phi\cdot\nabla\phi \right.
 \labell{act12}\\
&&\quad+ \cala_1\left(\phi\right) R^2+ \cala_2\left(\phi\right)
R_{ab}R^{ab} +\cala_3\left(\phi\right) R_{abcd}R^{abcd}
 \nonumber\\
&&\quad+\calb_{1}\left(\phi\right)\,\nabla\phi\cdot\nabla\phi \,R+
\calb_{2}\left(\phi\right)\,\jsquare\phi\, R
+\calb_{3}\left(\phi\right) \nabla^a\phi\nabla^b\phi \,R_{ab}
 \nonumber\\
&&\left.\quad+\calc_{1}\left(\phi\right)
(\nabla\phi\cdot\nabla\phi)^2 +
\calc_{2}\left(\phi\right)\,(\jsquare\phi)^2 +
\calc_{3}\left(\phi\right) \nabla\phi\cdot\nabla\phi\, \jsquare\phi
 \right]\,.\nonumber
 \eeqa
Now consider making field redefinitions: $g_{ab}\rightarrow
g_{ab}+\delta g_{ab}$ and $\phi\rightarrow \phi+\delta\phi$. The
most general field redefinition involving two-derivative
contributions can be written
 \beqa
\delta g_{ab} &=&
\calm_1\,R_{ab}+\calm_2\,\nabla_a\nabla_b\phi+\calm_3 \,
\nabla_a\phi\nabla_b\phi
 \labell{newer1}\\
&&+\left(\calm_4 R
+\calm_5\,\jsquare\phi+\calm_6\,\nabla\phi\cdot\nabla\phi +
\calm_7\right)\,g_{ab}\,,
 \nonumber\\
\delta \phi &=& \caln_1 R +\caln_2\,\jsquare\phi+\caln_3
\,\nabla\phi\cdot\nabla\phi\,.
 \nonumber
 \eeqa
In these expressions, all of the $\calm_i$ and $\caln_i$ are
understood to be functions of $\phi$ which are nonsingular at
$\phi=0$ and they are of order $\lp^2$. With these field
redefinitions, the leading change in the action is
 \beqa
\delta I&=&\frac{1}{2\lp^3}\int d^5x \sqrt{-g} \left\lbrace\left[
\frac{1}{2}\left(\calu(\phi)+R-\calk(\phi)\nabla\phi
\cdot\nabla\phi\right)g^{ab}-R^{ab}+\calk(\phi)\nabla^a\phi
\nabla^b\phi\right]\,\delta g_{ab}\right.
 \nonumber\\
&&\left.\vphantom{\frac{1}{2}}\qquad\qquad\qquad\qquad
+\biggl(\calu'(\phi)-2\calk(\phi)\jsquare\phi-\calk'(\phi)
\nabla\phi \cdot\nabla\phi\biggr)\,\delta\phi\right\rbrace
 \labell{change1}\\
&=&\frac{1}{2\lp^3}\int d^5x \sqrt{-g}\, \frac{1}{2}
\left\lbrace\vphantom{\frac{1}{2}} 5\calm_7\,\calu
+\left((\calm_1+5\calm_4)\calu+\calm_7\right)\,R
 \right.\nonumber\\
&&+\left((\calm_3+5\calm_6)\calu+3\calm_7\calk-\left[(\calm_2+5
\calm_5)\calu\right]'\right)(\nabla\phi)^2
 \nonumber\\
&&+\left(\calm_1+3\calm_4\right)R^2-\calm_1\,R_{ab}R^{ab}
+\left(\calm_2+3\calm_5-4\caln_1\calk\right)\jsquare\phi\,R
 \nonumber\\
&&+\left(\calm_3+3\calm_6+(\calm_1+3\calm_4)\calk\right)
(\nabla\phi)^2R-2\left(\calm_2-\calm_1\calk\right)
\nabla^a\phi\nabla^b\phi\, R_{ab}
 \nonumber\\
&&+\left(\left[\calm_2\calk\right]'-(\calm_3-3\calm_6)\calk
-2\caln_3\calk'\right)(\nabla\phi\cdot\nabla\phi)^2-4\caln_2\calk\,(\jsquare\phi)^2
 \nonumber\\
&&\left.\vphantom{\frac{1}{2}}\qquad\qquad
+\left((2\calm_2+3\calm_5-4\caln_3)\calk-2\caln_2\calk'\right)\,
(\nabla\phi)^2\jsquare\phi\right\rbrace\,,
 \labell{change2}
 \eeqa
where as above, the prime indicates a derivative with respect to
$\phi$. Note that we have integrated by parts to produce the
expressions in \reef{change2}. Now given this result is should be
clear that we have more than enough freedom to eliminate all of the
four-derivative scalar terms in \reef{act12}, \ie we can set to zero
the coefficients $\cala_{1,2}$, $\calb_{1,2,3}$ and $\calc_{1,2,3}$.
While we do not present the precise choices needed to produce these
cancellations, we note the various couplings in \reef{act12} can be
eliminated by fixing in turn various coefficients appearing in the
field redefinitions \reef{newer1}, as follows: $(\cala_1,\calm_4)$,
$(\cala_2,\calm_1)$, $(\calb_1,\calm_3)$, $(\calb_2,\caln_1)$,
$(\calb_3,\calm_2)$, $(\calc_1,\caln_3)$, $(\calc_2,\caln_2)$,
$(\calc_3,\calm_5)$. This leaves $\calm_6$ and $\calm_7$
undetermined. We can use the freedom in $\calm_7$ to prevent any
scalar couplings appearing in the Einstein term after the field
redefinition and to keep the Planck scale fixed. Hence the field
redefinitions \reef{newer1}, as well as integrating by parts, allow
us to simplify the general action \reef{act11} down to
 \beq
 I=\frac{1}{2\lp^3}\int d^5x
\sqrt{-g}\left[\,\calu\left(\phi\right) + R
-\calk(\phi)\,\nabla\phi\cdot\nabla\phi +\cala_3\left(\phi\right)
R_{abcd}R^{abcd}\right]\,.
 \labell{act33}
 \eeq
Given this result, it is clear that none of the higher order terms
involving derivatives of the scalar can be relevant in calculating
quantities such as the shear viscosity.

Unfortunately, it turns out that field redefinitions are not as
effective in eliminating four-derivative interactions when the
effective theory involves many scalars $\phi^k$. In this case, the
coefficients of each of the scalar field interactions in
\reef{act11} become ``tensors'' with indices to describe the various
independent interactions involving different combinations of
scalars. Hence the general four-derivative action for gravity
coupled to a set of scalar fields $\phi^k$ becomes:
 \beqa
 I&=&\frac{1}{2\lp^3}\int d^5x
\sqrt{-g}\left[\,\calu\left(\phi^m\right) + R
-\calk_{ij}(\phi^m)\,\nabla\phi^i\cdot\nabla\phi^j \right.
 \labell{act01}\\
&&\quad+ \cala_1\left(\phi^m\right) R^2+ \cala_2\left(\phi^m\right)
R_{ab}R^{ab} +\cala_3\left(\phi^m\right) R_{abcd}R^{abcd}
 \nonumber\\
&&\quad+
\calb_{1ij}\left(\phi^m\right)\,\nabla\phi^i\cdot\nabla\phi^j\,
R+\calb_{2i}\left(\phi^m\right)
\jsquare\phi^i\,R+\calb_{3ij}\left(\phi^m\right)
\nabla^a\phi^i\nabla^b\phi^j\,R_{ab}
  \nonumber\\
&&\quad +\calb_{4i}\left(\phi^m\right)
\nabla^a\nabla^b\phi^i\,R_{ab}+\calc_{1ijkl}\left(\phi^m\right)
\nabla\phi^i\cdot\nabla\phi^j\,\nabla\phi^k\cdot\nabla\phi^l +
\calc_{2ij}\left(\phi^m\right)\,\jsquare\phi^i\,\jsquare\phi^j
  \nonumber\\
&&\quad+ \calc_{3ijk}\left(\phi^m\right)
\nabla\phi^i\cdot\nabla\phi^j\, \jsquare\phi^k
+\calc_{4i}\left(\phi^m\right) \jsquare{}^2\phi^i
+\calc_{5ij}\left(\phi^m\right)
\nabla_a\phi^i\,\nabla^a\jsquare\phi^j
 \nonumber\\
&&\left. \quad +\calc_{6ij}\left(\phi^m\right)
\nabla_a\nabla_b\phi^i\,\nabla^a\nabla^b\phi^j
+\calc_{7ijk}\left(\phi^m\right)
\nabla_a\nabla_b\phi^i\,\nabla^a\phi^j\nabla^b\phi^k\right]\,.
 \nonumber
 \eeqa
Again, many of the four-derivative terms can be eliminated by simply
integrating by parts. However, there is one complication in
considering $\calc_{7ijk}$. The natural extension of \reef{examp1}
now comes from considering the following total derivative:
 \beqa
&&\nabla_a\left( \calc_{7(ij)k}\left(\phi^m\right)
\nabla\phi^i\cdot\nabla\phi^j\,\nabla^a\phi^k\right)=2\,
\calc_{7(ij)k}\left(\phi^m\right)
\nabla_a\nabla_b\phi^i\,\nabla^b\phi^j\,\nabla^a\phi^k
 \labell{totder}\\
&&\qquad\quad+\, \calc_{7(ij)k}\left(\phi^m\right)
\nabla\phi^i\cdot\nabla\phi^j\, \jsquare\phi^k
+\partial_l\calc_{7(ij)k}\left(\phi^m\right)
\nabla\phi^i\cdot\nabla\phi^j\,\nabla\phi^k\cdot\nabla\phi^l\,.
 \nonumber
 \eeqa
Above, the parentheses on the subscripts indicate symmetrization of
the indices, \ie
$\calc_{7(ij)k}=\frac{1}{2}\left(\calc_{7ijk}+\calc_{7jik}\right)$.
In general then, the coefficients $\calc_{7ijk}$ do not have to be
symmetric in the indices $i$ and $j$ but because of the form of the
tensor in the total derivative \reef{totder}, integrating by parts
can only eliminate the symmetric combination $\calc_{7(ij)k}$. Hence
the natural generalization of \reef{act12} is slightly more involved
in the case of multiple scalars. First we must add indices as
appropriate in the interactions appearing there but we must also
include an extra term proportional to $\calc_{7[ij]k} =\frac{1}{2}
\left(\calc_{7ijk}-\calc_{7jik}\right)$.

Next we wish to consider the field redefinitions generalizing those
in \reef{newer1}, \ie $g_{ab}\rightarrow g_{ab}+\delta g_{ab}$ and
$\phi^i\rightarrow \phi^i+\delta\phi^i$ with
 \beqa
\delta g_{ab} &=&
\calm_1\,R_{ab}+\calm_{2i}\,\nabla_a\nabla_b\phi^i+\calm_{3ij} \,
\nabla_a\phi^i\,\nabla_b\phi^j
 \labell{newer2}\\
&&+\left(\calm_4 R
+\calm_{5i}\jsquare\phi^i+\calm_{6ij}\nabla\phi^i\cdot\nabla\phi^j +
\calm_7\right)\,g_{ab}\,,
 \nonumber\\
\delta \phi^i &=& \caln_1^i R
+\caln_{2j}^i\jsquare\phi^j+\caln_{3jk}^i\,
\nabla\phi^j\cdot\nabla\phi^k\,.
 \nonumber
 \eeqa
With these field redefinitions, we can consider the leading change
in the action but this exercise is rather tedious and so we give
only a schematic description of the results. In certain cases, the
previous discussion follows through unchanged. For example above, we
canceled $\calb_1$ by fixing the coefficient $\calm_3$. Here this
pairing becomes $(\calb_{1(ij)} ,\calm_{3(ij)})$. The structure of
the corresponding terms is such that both of these expressions are
symmetric in their subscripts, as indicated by the parentheses.
Hence the index or tensor properties match nicely in this particular
case and it is clear that there are precisely enough degrees of
freedom in $\calm_{3(ij)}$ to eliminate $\calb_{1(ij)}$. However, in
a number of cases, there is a mismatch for the tensor expressions.
For example, the pairing $(\calc_{3},\calm_{5})$ becomes with
multiple scalars, $(\calc_{3(ij)k}, \calm_{5i})$. Hence in this
case, it is clear that in general with more than one scalar field,
there are not enough degrees of freedom in the field redefinition
$\calm_{5i}$ to eliminate all of the possible couplings
$\calc_{3(ij)k}$. Our final result is that we still have the freedom
to set to zero the couplings, $\cala_{1}$, $\cala_{2}$,
$\calb_{1(ij)}$, $\calb_{2i}$ and $\calc_{2(ij)}$, but we can only
partially eliminate $\calb_{3(ij)}$, $\calc_{1(ij)(kl)}$ and
$\calc_{3(ij)k}$. Further as described above, a new set of couplings
arise from $\calc_{7[ij]k}$. Of course, in any given theory, it may
be that the full set of general couplings does not appear, \eg there
might be internal symmetries which restrict the number and form of
the independent couplings.

Hence after making the $\calo(\lp^2)$ field redefinitions
\reef{newer2}, the general four-derivative action \reef{act01} can
be simplified to take the form
 \beqa
 I&=&\frac{1}{2\lp^3}\int d^5x
\sqrt{-g}\left[\,\calu\left(\phi^m\right) + R
-\calk_{ij}(\phi^k)\,\nabla\phi^i\cdot\nabla\phi^j +
\cala_3\left(\phi^m\right) R_{abcd}R^{abcd}\right.
 \labell{act033}\nonumber\\
&&\quad +\calb_{3(ij)}\left(\phi^m\right)
\nabla^a\phi^i\,\nabla^b\phi^j\,R_{ab}
+\calc_{1(ij)(kl)}\left(\phi^m\right) \nabla\phi^i\cdot\nabla\phi^j
\,\nabla\phi^k\cdot\nabla\phi^l
  \nonumber\\
&&\quad\left. + \calc_{3(ij)k}\left(\phi^m\right)
\nabla\phi^i\cdot\nabla\phi^j\, \jsquare\phi^k
+\calc_{7[ij]k}\left(\phi^m\right) \nabla_a\nabla_b\phi^i\,
\nabla^b\phi^j\,\nabla^a\phi^k\right]\,.
 \eeqa
At this point, an important observation is that the higher order
interactions in the second and third line of this action contain at
least two factors with derivatives of the scalars. Hence, since we
are treating these terms perturbatively, if the scalars are constant
in the leading solution, they will remain constant at the next
order. Certainly, the scalars are constant in the leading order
background of an AdS$_5$ black hole \reef{Ts} and so the scalars
will not effect the thermodynamic properties of dual CFT (at least
at this order in the expansion in $(c-a)/c$). That is, these new
coefficients define new parameters of the CFT which characterize
certain correlators of the new operators (dual to the scalars) and
the stress tensor. However, the properties of the thermal stress
tensor are independent of these parameters at this order.

This discussion can be further extended to include vectors in the
gravity theory. While a complete discussion requires an even more
elaborate analysis, it is relatively straightforward to show that
any new four-derivative interactions are at least quadratic in the
field strengths of the gauge fields, with one exception. Hence an
argument similar to that above applies here as well, with the
conclusion that these terms will not effect the CFT's thermal
properties, at this order. The one exception to these statements is
as follows: In five dimensions with a U(1) gauge field, we can
introduce an interaction: $\int A\wedge R^a{}_b\wedge R^b{}_a$. This
term plays an interesting role in describing the anomaly for the
$U(1)_R$ current \cite{anom,anom2} -- see also \cite{sugra} for
recent supergravity analysis of this term. Since this interaction is
linear in the gauge potential, it will induce a nontrivial profile
in a background where $R^a{}_b\wedge R^b{}_a$ is nonvanishing.
However, this combination of curvatures vanishes both for the
AdS$_5$ vacuum and an AdS$_5$ black hole background. Hence we can
conclude again that this term will play no role in determining the
thermal properties of the CFT.

\section{String theory origin of $R^2$}\label{string}

In the string theory example considered by Kats and Petrov \cite{kp}
and more generally in the F-theory constructions of \cite{at}, the
curvature-squared interaction is argued to arise from the
world-volume action of the D7-branes \cite{at,anom2}. In this
appendix, we would like to develop a schematic understanding of the
parameter dependence of the coupling coefficients in these higher
derivative interactions. In particular, we contrast these couplings
with the analogous coefficients in the celebrated $R^4$ interaction
\cite{four} that arises from the closed string sector. Hence we are
able to confirm the conditions \reef{dom} under which the $R^2$
corrections coming from the branes dominate over the bulk
corrections arising the $R^4$ interaction. Along the way we will
motivate the usage of the $R^2$ terms in eq. (\ref{act3}). We must
note though that the final results rely on treating the D$p$-branes
as probe branes and so the discussion has more limited applicability
than the effective action approach in section \ref{dual}.
Schematically we can write
\begin{equation}
S=\kappa_1 \int d^{10}x \sqrt{-g} (R-F_5^2+\alpha'^3 R^4+\cdots)
-\kappa_2 \int d^{p+1} x (\sqrt{G+F}+\alpha'^2\sqrt{G}
R^2+\cdots)\,.
\end{equation}
We are in Einstein frame. Terms arising at $\ls^6$ in the bulk
action are generically denoted by $R^4$. By $R^2$ we mean a generic
term arising at $\ls^4$ order in the brane action. It is known from
scattering amplitude calculations off D-branes and O-planes
\cite{sign} that these terms arise as stringy corrections to the DBI
action. Here
\begin{equation}
\kappa_1\sim \frac{1}{g_s^2 \ls^8}\,,\qquad \kappa_2\sim\frac{N_f}{
g_s \ls^{p+1}}\,.
\end{equation}
$\kappa_1$ is essentially the inverse of the
ten-dimensional Newton's constant while $\kappa_2$ is related to the
tension of the D$p$-brane. We will consider $p>3$ so that we can get
a 5d action by wrapping the brane on some $p-4$ cycle. So $p=5,7,9$.

Now we know from the standard AdS/CFT dictionary that
\begin{equation}
\ls^2\sim \frac{1}{\sqrt{\lambda}}\,,\qquad g_s\sim
\frac{\lambda}{\nc}\,,
\end{equation}
where for convenience we have set the AdS radius to unity. Thus when
$\lambda$ is large, the six and higher derivative terms in the full
brane action can be ignored compared to the four-derivative terms.
Then in terms of these variables
\begin{eqnarray}
S&\sim&{\nc^2}\bigg(\int d^{10}x \sqrt{-g} (R-F_5^2+ \frac{1}{
\lambda^{3/2}} R^4+\cdots)-N_f\int d^{p+1}x \frac{1}{ \nc
\lambda^{(3-p)/4}} \sqrt{G+F} \nonumber \\&+& \frac{1}{ \nc
\lambda^{(7-p)/4}} \sqrt{G}R^2+\cdots\bigg)\,.
\end{eqnarray}

Thus for $p=7$, the first term in the DBI action leads to a
correction to the effective cosmological constant of
$\calo(\lambda/\nc)$ while the second term gives an $R^2$ term with
coefficient $\calo(1/\nc)$. In order to produce a 5d theory, we need
to integrate the brane action over some ($p$--4)-cycle. We have to
ensure that the volume of this ($p$--4)-cycle satisfies $V_{p-4}\gg
\ls^{3}$ for the derivative expansion to make sense. For general
$p$, from the 5d point of view we have
\begin{equation}
\begin{split}
S_5\sim &\nc^2 V_5 \int d^{5}x \sqrt{-g_5} \biggl( R_5-2\Lambda
+\frac{1}{ \lambda^{3/2}} R_5^4+\cdots- \frac{V_{p-4}}{V_5}\frac{N_f
}{\nc }\,\lambda^{(p-3)/4}
\\
&-\frac{V_{p-4}}{V_5} \frac{N_f}{\nc}\, \lambda^{(p-7)/4}
R_5^2+\cdots\biggr)\,.
\end{split}
\end{equation}
In imposing various constraints, first we require
\begin{equation}
V_5\gg\frac{1}{ \lambda^{5/4}}\,, \qquad{V_{p-4}}\gg \frac{1}{
\lambda^{(p-4)/4}}\,, \qquad \lambda \gg 1\,,\qquad \,.
\end{equation}
for the derivative expansion to be sensible. Next from the above
action, we see that for the $R^2$ terms to produce the leading
curvature corrections, we must have
\begin{equation}
1\gg\frac{V_{p-4}}{V_5}\frac{N_f}{ \nc}\,\lambda^{(p-7)/4} \gg
\frac{1}{ \lambda^{3/2}}\,, \quad {\rm or~~}
\lambda^{(7-p)/4}\gg\frac{V_{p-4}}{V_5}\frac{N_f}{\nc}\gg
\lambda^{(1-p)/4}\,.\label{goat}
\end{equation}
We also require that the brane tension does not produce a large
modification to the cosmological constant (\eg change the sign of
$\Lambda$) which gives
\begin{equation}
\lambda^{(3-p)/4}\gg\frac{V_{p-4}}{V_5}\frac{ N_f }{ \nc }\,.
\end{equation}
The latter replaces the first inequality in \reef{goat} giving a
more stringent constraint. The second inequality in \reef{goat}
yields
\begin{equation}
\lambda\gg \left(\frac{V_5}{V_{p-4}} \frac{\nc}{
N_f}\right)^{\frac{4}{ p-1}}\qquad \Longrightarrow \qquad
\frac{\lambda}{\nc}\gg \left(\frac{V_5}{ V_{p-4}}\frac{1}{N_f}
\right)^{\frac{4}{ p-1}} \nc^{\frac{5-p}{ p-1}}\,.\labell{goat1}
\end{equation}
If only string loop effects, unsuppressed by powers of $\ls$, were
present then these would dominate if $p\leq 9$.  Thankfully,
supersymmetry prevents string loop corrections to the lowest order
DBI terms and hence this situation does not arise \cite{tseytlin}.

Finally this formal analysis treats the back-reaction of the
D-branes perturbatively and so we must insist on weak string
coupling, $g_s\ll 1$. Hence we require that
 \beq \frac{\lambda}{\nc}\ll 1 \labell{goat2}
 \eeq
Comparing to \reef{goat1}, we must then have $p>5$. In other words,
only for $p=7,9$ can the $R^2$ term be viewed sensibly as arising
from a probe brane and dominating the $\ls^6$ terms. Further,
however, the D9 brane case is not viable since the zero-temperature
limit is not supersymmetric. Thus it appears that we can only use
D7-branes as sources for the $R^2$ term in a perturbative setting.
Typically there are also couplings of the type
\begin{equation}
T_p \int C \wedge e^F \wedge [{\rm tr}(R_T\wedge R_T)-{\rm tr}
(R_N\wedge R_N)]\,,
\end{equation}
where $N$ and $T$ denote the normal and tangent bundles
respectively. One might worry that such couplings would change $C$
and $\calo(\ls^4)$ and hence feedback at the same order in the
Einstein equations. Fortunately for the case of interest this does
not happen as can be explicitly checked. The modification only
occurs at $\calo(\ls^8)$ which can be ignored. In 5d-language this
translates into ignoring $\ls^4$ modifications to $A_\mu$.

The next important question to fix is the sign of the $R^2$ term.
Following the general strategy discussed in section \ref{dual}, we
can construct a effective theory in five dimensions for which the
coefficient of the $R^2$ correction is fixed by the trace anomaly of
the gauge theory. As in \cite{bng}, flux terms arising at this order
will be ignored. It is interesting then that scattering amplitudes
in string theory from D-branes and O-planes seem to indicate that
the sign is positive \cite{sign}.\footnote{In \cite{panda}, it is
shown that supersymmetry leads to a positive coefficient for
$R_{abcd}R^{abcd}$ also in the case of heterotic strings.}

\end{document}